\documentclass[preprint,12pt]{elsarticle}

\usepackage{booktabs}
\usepackage{subfigure}
\usepackage{pgfplots}
\pgfplotsset{width=5.5cm}
\AtBeginDocument{%
  \providecommand\BibTeX{{%
    \normalfont B\kern-0.5em{\scshape i\kern-0.25em b}\kern-0.8em\TeX}}}

\usepackage{amssymb}
\usepackage{balance}  
\usepackage{graphics} 
\usepackage{color}
\usepackage{booktabs}
\usepackage{ccicons}

\usepackage{url}
\usepackage{fancybox}
\usepackage{multirow}
\usepackage{flushend}
\usepackage{booktabs}
\usepackage{tabularx}
\usepackage{comment}
\usepackage{array}
\usepackage[flushleft]{threeparttable}
\usepackage{mdframed}
\graphicspath{{}{images/}{dia/}}
\DeclareGraphicsExtensions{.pdf,.png,.svg}

\usepackage{listings}
\usepackage{courier}
\usepackage{hyperref}

\usepackage{xpatch}

\xpatchcmd{\refstepcounter}{%
  \stepcounter{#1}%
}{%
  \stepcounter{#1}%
  \xdef\scr{\number\value{#1}}%
}{\typeout{success}}{\typeout{failure}}

\newcounter{o}
\usepackage{tikz}
\definecolor{1c1}{RGB}{188,162,6}
\definecolor{1c2}{RGB}{137,129,80}
\definecolor{1c3}{RGB}{239,167,31}
\definecolor{1c4}{RGB}{88,194,241}
\definecolor{1c5}{RGB}{6,180,188}

\tikzset{mynode/.style={draw=white,solid,circle,fill=green,inner sep=1pt, thick,
text=black}}
\tikzset{arrow line/.style={dashed, line width= 2.5pt, color=#1}}
\usepackage{varwidth}
\def\bf{\textbf}

\def\fig {Figure~}

\def\tbl {Table~}

\def\sec {Section~}
\def\secs {Sections~}

\def\it{\textit}

\def\tt{\mct}

\newcommand{\mct}[1]{{\footnotesize {\texttt {#1}}}}
\newcommand{\rev}[1]{\textcolor{blue}{ #1}}

\usepackage{paralist}

\usepackage[figuresright]{rotating}
\newcommand{\nd}{\vspace{1mm}\noindent}
\usepackage{tikz}

\lstdefinestyle{inlinecode}{basicstyle={\ttfamily\scriptsize\bfseries}}

\newcommand{\urls}[1]{{\scriptsize\url{#1}}}
\usepackage{tcolorbox}
\newcommand{\emt}[1]{\emph{``#1''}}

\usepackage{paralist}
\usepackage[outercaption]{sidecap}
\usepackage [autostyle, english = american]{csquotes}
\MakeOuterQuote{"}
\newcounter{scn}
\usepackage[shortlabels]{enumitem}
\usepackage{tikz}
\usepackage{styles/pgf-pie}
\usepackage{paralist}
\usepackage{soul}

\usetikzlibrary{positioning,shadows}
\usepackage{balance}
\newcommand{\dq}[1]{\href{https://stackoverflow.com/questions/#1/}{$Q_{#1}$}}

\newif\ifpienumberinlegend
\pgfkeys{/number in legend/.code=
    \expandafter\let\expandafter\ifpienumberinlegend
    \csname if#1\endcsname
    \ifpienumberinlegend

    \def\beforenumber##1\afternumber{}%
    \fi,
    /number in legend/.default=true
}
\definecolor{1c1}{RGB}{188,162,6}
\definecolor{1c2}{RGB}{137,129,80}
\definecolor{1c3}{RGB}{239,167,31}
\definecolor{1c4}{RGB}{88,194,241}
\definecolor{1c5}{RGB}{6,180,188}

\tikzset{mynode/.style={draw=white,solid,circle,fill=green,inner sep=1pt, thick,
text=black}}
\tikzset{arrow line/.style={dashed, line width= 2.5pt, color=#1}}

\usepackage{bchart}

\usepackage{varwidth}					
\usepackage{adjustbox}

\journal{Journal IST}

\begin{document}

\begin{frontmatter}


\title{An Empirical Study of IoT Security Aspects at Sentence-Level in Developer Textual Discussions}



\author{Nibir Mandal and Gias Uddin}

\address{DISA Lab, University of Calgary, Canada.}

\begin{abstract}
\bf{Context:} IoT is a rapidly emerging paradigm that now encompasses almost every aspect of our modern life. As such, ensuring the security of IoT devices is crucial. IoT devices can differ from traditional computing (e.g., low power, storage, computing), thereby the design and implementation of proper security measures can be challenging in IoT devices. We observed that IoT developers discuss their security-related challenges in developer forums like Stack Overflow (SO). However, we find that IoT security discussions can also be buried inside non-security discussions in SO.

\nd\bf{Objective:} In this paper, we aim to understand the challenges IoT developers face while applying security practices and techniques to IoT devices. We have two goals: \begin{inparaenum}[(1)]
\item Develop a model that can automatically find security-related IoT discussions in SO, and 
\item Study the model output (i.e., the security discussions) to learn about IoT developer security-related challenges.
\end{inparaenum}

\nd\bf{Method:}  First, we download all 53K posts from StackOverflow (SO) that contain discussions about various IoT devices, tools, and techniques. Second, we manually labeled 5,919 sentences from 53K posts as 1 or 0 (i.e., whether they contain a security aspect or not). Third, we then use this benchmark to investigate a suite of deep learning transformer models. The best performing model is called SecBot. Fourth, we apply SecBot on the entire 53K posts and find around 30K sentences labeled as security. Fifth, we apply topic modeling to the 30K security-related sentences labeled by SecBot. Then we label and categorize the topics. Sixth, we analyze the evolution of the topics in SO.

\nd\bf{Results:} We found that 
\begin{inparaenum}[(1)]
\item SecBot is based on the retraining of the deep learning model RoBERTa. SecBot offers the best F1-Score of .935,
\item there are six error categories in misclassified samples by SecBot. SecBot was mostly wrong when the keywords/contexts were ambiguous (e.g., `gateway' can be a security gateway or a simple gateway),
\item there are 9 security topics grouped into three categories: Software, Hardware, and Network, and
\item the highest number of topics belongs to software security, followed by network security and hardware security.
\end{inparaenum}

\nd\bf{Conclusion:} IoT researchers and vendors can use SecBot to collect and analyze security-related discussions from developer discussions in SO. The analysis of nine security-related topics can guide major IoT stakeholders like IoT Security Enthusiasts, Developers, Vendors, Educators, and Researchers in the rapidly emerging IoT ecosystems.
\end{abstract}

\begin{keyword}
IoT, security, Stack Overflow, Deep Learning, Empirical Study
\end{keyword}

\end{frontmatter}



\section{Introduction}\label{sec:intro}
Internet of Things (IoT) is a rapidly emerging paradigm that is defined as the connection between places and physical objects (i.e., things) over the Internet via smart computing 
devices~\cite{Atzori-SurveyIoT-ComputerNetwork2010,Gubbi-IoTVisionDirection-FGCS2013}. This technological revolution now encompasses almost 
every aspect of our modern life, such as smart homes, cars, voice-enabled intelligent devices, and so on~\cite{Fuqaha-IoTSurveyTechnologiesApplications-IEEECST2015,Pretz-TheNextEvolutionInternet-IEEEMagazie2013}.
Indeed, according to Statistica reports~\cite{iot-numbers}, the number of ``smart'' connected devices 
was 15 billion in 2015 and it is projected to be 75 billion by 2025 (i.e., 400\% increase in ten years). As such, it is not surprising that interest 
in the IoT technologies is growing among developers to develop new architectures, tools, and techniques and to make those devices intelligent based on data analytics~\cite{Marjani-IoTDataAnalytics-IEEEAccess2017,Weyrich-RefArchitectureIoT-IEEESoftware2016}.

The pervasiveness of IoT-based solutions in our daily lives has raised concerns about security 
in IoT~\cite{Chi-SmartHomeCrossAppInference-DSN2020,Ding-IoTSafetyPhysicalInterfaction-CCS2018,Edwards-IoTHajimeWorm-Rapidly2016}. 
Such concerns for IoT devices
can be related to the implementation/adoption of security protocols (e.g., zigbee chain reaction~\cite{Ronen-IoTNuclearZigbeeChainReaction-SP2017}) and 
roles (e.g., authentication~\cite{Gong-IoTPIANO-ICDCS2017}), communication over the network (e.g., cross-device inference~\cite{Chi-SmartHomeCrossAppInference-DSN2020}), 
as well as the underlying hardware~\cite{Ho-IoTSmartLocks-ASIACCS2016}. Indeed, ensuring the security of billions/trillions of IoT devices requires efforts of an unprecedented nature, 
unlike anything we have seen before~\cite{Sekar-HandlingTrillionIoTSecurity-HotNets2015}. As such, it is important to understand 
the challenges IoT developers face during their adoption of security principles and practices for IoT devices so that we can 
design effective techniques to address these challenges. 
 
With interest in IoT growing, we observe discussions of IoT developers in online forums like Stack Overflow (SO). 
SO is a large online community where millions of developers ask and
answer questions. To date, there are
around 120 million posts and 12 million registered users on SO~\cite{website:stackoverflow}. 
Previously, several research has been conducted to
analyze SO posts, e.g., to analyze discussions on big
data~\cite{Bagherzadeh2019}, concurrency~\cite{Ahmed-ConcurrencyTopic-ESEM2018},
programming issues~\cite{Barua-StackoverflowTopics-ESE2012}, blockchain
development~\cite{Wan-BlockChainTopicSO-IEEETSE2019}, microservices~\cite{bandeira2019we}, and
security~\cite{yang2016security}. However, we are aware of no
research that analyzed IoT security-related discussions on SO, although such insight can complement existing IoT literature, which so far has mainly used surveys 
to understand the issues and needs of IoT practitioners~\cite{Atzori-SurveyIoT-ComputerNetwork2010,Gubbi-IoTVisionDirection-FGCS2013,Fuqaha-IoTSurveyTechnologiesApplications-IEEECST2015}.

We observe that IoT developers are active in SO and they discuss the pros and cons of any IoT tools, devices, and technologies. Moreover, they are aware of the new release of IoT devices and there are sufficient discussions to get a full overview of security details in IoT devices. For example, a developer asked about `arduino' security details in SO question \dq{48854031}. The discussion session contains multiple security concerns for `arduino' where well-known developers share their valuable knowledge, including implementation support. This discussion is helpful for both new and advanced developers as they get insightful information and various aspects of the IoT\cite{iot-discussion}. Besides this, some recent studies indicate that they are currently facing challenges while developing or implementing IoT devices due to the intricacies and sensitivity of security details \cite{iot-security-challenges}. Thus, the developers require proper knowledge and standard implementation details to develop a secure device or tools. As IoT developers are following SO and sharing knowledge with others subtly, the community can follow the statutes available in SO while developing the security details of an IoT product. Moreover, the community can form a standard for developing IoT tools, which they can propagate through SO discussions. This will be helpful for the community to build a secure device and the users to have a secure device. In addition, the newcomer can gather IoT security knowledge from SO, which will help them to build more robust devices in the future, as the security standard will be continuously upgraded in the future. This will be beneficial for the vendors as well, because they will get direct feedback about their product from the community, and this will help them to keep up with the market demands. Therefore, the IoT security discussion in SO has great value for IoT stakeholders.

\begin{figure}[t]
\centering
   	\includegraphics[scale=.75]{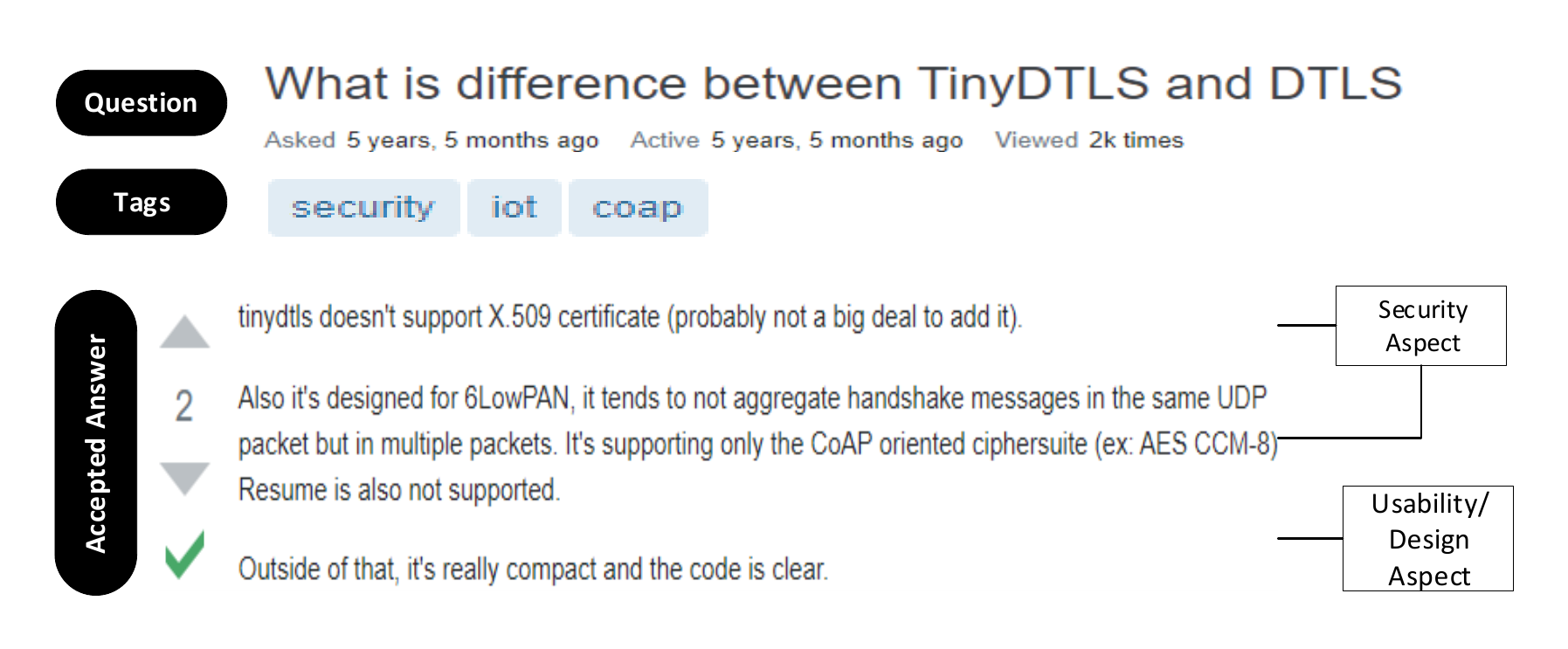}
   	\caption{An IoT discussion in a Stack Overflow answer about TinyDTLS security and usability aspects (\dq{33004863})}
   	 \label{fig:motivation-post-sentence}
\vspace{-5mm}
\end{figure}
A major problem in supporting the automated analysis of IoT security discussions from SO is that such security discussions can be buried inside 
technical discussions of developers. For example, consider the answer to the SO question \dq{33004863} in \fig\ref{fig:motivation-post-sentence}. 
The question concerns two security software libraries (i.e., APIs), both of which contain the implementation of the `Datagram Transport Layer Security (DLTS)' protocol. 
The DLTS is a communication protocol to provide security for software applications that operate in packet-switched networks where a datagram is a basic transfer unit. 
A datagram has a header and payload section. The protocol uses datagrams in a stateless manner to prevent malicious attacks like eavesdropping, tampering, or message 
forgery. The asker of \dq{33004863} is particularly interested in the DLTS implementation in the tinyDLTS library because he needs to use the protocol in his IoT device. 
However, he is also worried about whether the tinyDLTS API offers limited features due to constrained resources and data size. The answer to the question does confirm that 
the tinyDLTS has limited security features, e.g., it does not support X.509 certificates. The last line of the answer, however, notes that the API is usable. 
Indeed, while the first three sentences 
in the answer are related to the security features of the tinyDLTS API, the last line is about the usability aspect of the API. Therefore, if we simply 
are interested in learning about security aspects of software in SO posts, we may want to only consider the first three sentences of the answer. This means 
that sentence-level security aspect detection in SO posts can be more precise and less noisy than post-level analysis. 
Therefore, we need automated techniques to precisely detect security aspects in SO developer discussions. 
Previously, Uddin and Khomh~\cite{Uddin-OpinionValue-TSE2019} developed a supervised shallow machine learning classifier to detect security aspects at sentence-level in 
SO posts. However, the classifier was not trained and tested on IoT-related sentences. This can affect the performance of the classifier model as IoT security is different from traditional security. For example, in the IoT domain, X.509 indicates a public key certificate that is related to security. But in our traditional computation, we can consider this an entity or a number. A generic classifier model can't discern such security details from non-security discussions as the model is trained and designed to have generic knowledge. In fact, a security-specific model can not discern such discussions. Thus, we need to design a classifier model that has IoT domain knowledge. Moreover, the classifier that Uddin et al.\cite{Uddin-OpinerReviewAlgo-ASE2017} used also did not use recent advances in deep learning transformer models, which are found to outperform shallow learning models in software engineering (SE) datasets (e.g., sentiment classification~\cite{Zhang-SentimentDetectionSEBERT-ICSME2020}).
As a first step towards developing techniques to support the detection and analysis of IoT security discussions in SO at sentence-level, in this paper, we adopt two phases. 
\begin{itemize}
    \item \bf{Phase 1.} We studied the effectiveness of a suite of 
deep learning (DL) models to detect security aspects on IoT discussions in SO.
    \item \bf{Phase 2.}  We studied the IoT topics found in the security-related sentences in SO, where the security-related sentences are detected by the best performing model from Phase 1.
\end{itemize}

\bf{\ul{In Phase 1}}, our focus is to study and find the best performing machine learning (ML) model that we can use to automatically detect security-related IoT discussions in SO. Detection of security aspects is the key to study IoT security-related discussions in SO. This is because, to the best of our knowledge, there is no existing technique to identify the security aspect of IoT posts. So, We collect 53K posts from SO based on 78 IoT tags (e.g., Arduino, Raspberry-pi). Then, we create an IoT security dataset of 5919 sentences from the IoT posts to assess the performance of deep learning (DL) models. In phase 1, we answered a total of three research questions.

\nd\bf{RQ$_1$. How  do  the  deep  learning  models  perform  to  detect  IoT  security aspects in the benchmark?}

 We studied four optimized transformers- i.e., BERT\cite{Delvin-BERTArch-Arxiv2018}, RoBERTa\cite{Liu-Roberta-Arxiv2019}, XLNet\cite{Yang-Xlnet-Arxiv2020}, and DistilBERT\cite{Distil-Bert} for this research question. Transformer-based advanced language models \cite{transformers-kant} are found to outperform other models in text classification (e.g., sentiment detection in SE~\cite{Zhang-SentimentDetectionSEBERT-ICSME2020}). In addition, recent studies suggest that a domain-specific DL model is more effective than a general-purpose deep learning model \cite{domain-specific-bert-tai}. Thus, for this research question, we also employ the BERTOverflow model, which has been pre-trained on SO dumps~\cite{tabassum2020code}. Indeed, in our security aspect detection, we find that the transformer models outperform the baseline SVM, which was found to be the most effective in security aspect detection by Uddin and Khomh~\cite{Uddin-OpinerReviewAlgo-ASE2017} in a general purpose API review dataset. RoBERTa shows the best performance among the four transformer models, with an F1-Score of 0.935. We name this trained model SecBot.

\nd\bf{RQ$_2$. What are the error categories in the misclassified cases of the best performing model?}

From the model's perspective, we analyze the misclassification of SecBot. It helps us get insights into the model and figure out the challenges of detecting security aspects in IoT discussions. We find that out of 5,919 samples in the IoT security dataset, SecBot misclassifies 139 sentences. We will analyze all 139 sentences. We discover six types of erroneous categories (i.e., `Ambiguous Keywords', `Ambiguous Context', `Implicit Context', `Code Parsing Error', `Text Parsing Error', and `Infrequent/unknown Keywords') in those samples.

\bf{\ul{In Phase 2}}, we apply SecBot to all sentences in our 53K IoT posts to detect security-related sentences. Out of the total 672K sentences, SecBot found around 30K security-related sentences. We studied the topics developers discussed about IoT security by answering two research questions.

\nd\bf{RQ$_3$. What  are  topics  in  the  IoT  security  discussions  collected  from  the entire IoT SO dataset using the best performing model?}

Topic modeling excavates useful information in a large dataset. In our case, we can learn about the security topics that developers ask, post, and answer in SO. Using SecBot, we found around 30K security-specific sentences. We apply topic modeling to the 30K sentences to learn about IoT security topics in developer discussions. We find 9 topics grouped into three categories: Software, Network, and Hardware. The largest number of topics belong to the software category, followed by the network category and the hardware category.

\nd\bf{RQ$_4$. How do the IoT security topics evolve in SO?}

Evolution studies of IoT security topics can be helpful for both developers, vendors, and investors to keep themselves up to date. IoT developers can learn about recent tools and technology where vendors and investors can learn about the developer's interests in any technology or tool. We find that while software has the greatest number of topics, in recent years since 2017, we see an almost equal number of security-related sentences for both network and software-related topics. This finding indicates that we need to equally prioritize IoT security research and development efforts for IoT software and networking related problems.


\nd\bf{Replication Package}: We share all the data and and all the source to replicate our study in \url{https://bit.ly/3nL2Sj5}.

\section{Empirical Study Setup}
In this section, we first offer a schematic overview of the overall empirical study approach (\sec\ref{sec:overview-study}).  In \sec\ref{sec:iot-data-collection}, we discuss the how we collect and process our IoT datasets from SO and in \sec\ref{iot-dataset-creation}, we discuss how we create our benchmark of 5919 sentences that we used to investigate the accuracy of our IoT security classification models in detecting IoT security-related sentences in SO. We then discuss how we fine-tune the five transformer models during our study (\sec\ref{sec:studied-models}) and offer details of the performance metrics used to report the accuracy of the models (\sec\ref{sec:performance-metrics}).
\begin{figure}[!t]
    \centering
    \includegraphics[scale=.28]{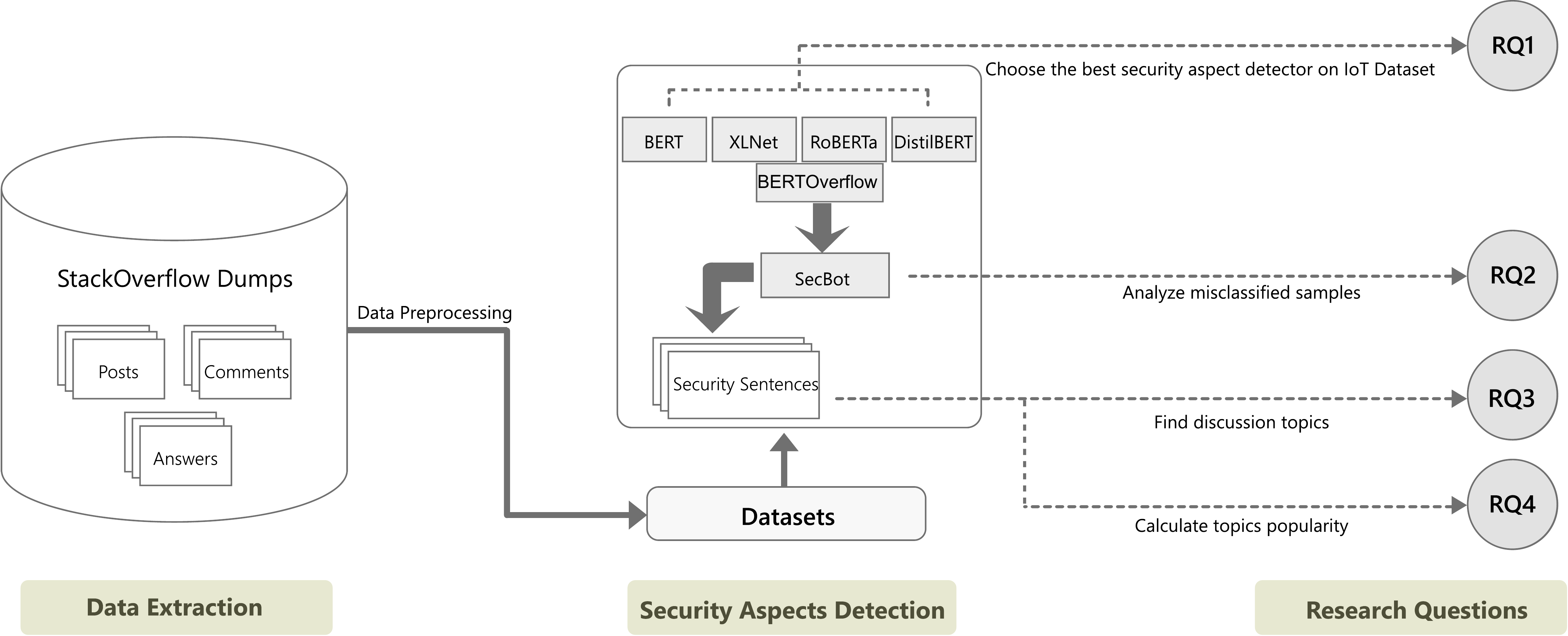}
    \caption{Overview of our empirical study}
    \label{fig:overview}
\end{figure}
\subsection{Schematic Overview of Study} \label{sec:overview-study}
In \fig \ref{fig:overview}, we show the overview of our empirical study and the formulation of our research questions. First, we collect IoT-relevant data (i.e., posts, comments, and answers) from SO dumps, applying some filters on SO-tags. Then we apply some preprocessing (i.e., sentence tokenization) to the dataset. To detect security-related sentences, we use five DL transformer models: BERT, RoBERTa, XLNet, DistilBERT, and BERTOverflow on the dataset. 
In RQ1, we evaluate five of the transformers (BERT, RoBERTa, XLNet, BERTOverflow, and DistillBERT) to find the best performer based on both domain-neutral and domain-specific word embedding. Our next objective is to understand the weaknesses of the security detectors of the best-performing model from RQ1. So we collect the predictions of the best security detectors. Then, we analyze the error categories of the best performing model from RQ1 in RQ2.
In RQ4, we apply topic modeling to find the most discussed topics and their distribution in the IoT security domain. Then we calculate the popularity of each topic category in RQ5 to understand the evolution of these topics.


\subsection{IoT Data Collection from Stack Overflow (SO)}\label{sec:iot-data-collection}
Our research started in 2020, when the latest SO data dump available was the September 2019 data dump from SO. Given that it is now 2022, newer SO data dumps are now available, e.g., the Jan 2022 dump. However, when we compared the SO dump of September 2019 with the SO data dump of Jan 2022, we observed several discrepancies like missing/deleted posts in the Jan 2022 data dump compared to the dump of September 2019. As such, in this paper, we analyzed both data dumps, i.e., September 2019 and January 2022. While we answer the research questions in our empirical study using the SO data dump of September 2019 in \sec\ref{sec:results}, we compared the major results from our empirical study between the two
SO data dumps (i.e., Sept 2019 vs Jan 2022) in \sec\ref{sec:discussion}. For both of the data dumps, our data collection process remains the same. For clarity, we describe our data collection process below by referring to SO data dump of September 2019.

We follow Uddin et al. \cite{uddin-iot} to collect SO posts related to IoT discussions. \it{\ul{First}}, we download the SO data dump~\cite{website:stackoverflow-datadump} 
of September 2019. The SO dataset includes all posts for 11 years between 2008 and September 2019. In total, it has 46,767,375 questions and answers. 
Out of those, around 40\% are questions and 60\% are answers. 
Out of the answers, around 34\% are accepted. \it{\ul{Second}}, we determine the list of potential IoT security tags in SO, following Yang et al.~\cite{yang2016security}. We identify three most popular IoT related tags, i.e., `iot', `arduino', and `raspberry-pi' in SO. As these three have the maximum number of posts in IoT domains, our intuition is that these tags should cover the widest range of security-related topics. Thus, we select these three tags as our primary tags. We denote those as $\mathcal{T}_{init}$. Then we collect candidate tags $\mathcal{T_A}$ related to $\mathcal{T}_{init}$ by analyzing tags that are labeled at least one of those tags $\mathcal{T}_{init}$. As this step includes all IoT tags, it makes sure that all security related tags are also added for the study. Following previous research~\cite{yang2016security,Bagherzadeh-BigdataTopic-FSE2019}, 
we systematically filter out \it{irrelevant} tags and 
finalize a list of all potential IoT tags $\mathcal{T}$ as follows. For each tag $t$ in $\mathcal{T_A}$, we compute its significance and relevance as follows.  
\begin{equation}
\textrm{Significance}~\mu(t) = \frac{\# \textrm{of Questions with tag $t$ in $\mathcal{P}$}}{\# \textrm{of Questions with tag $t$ in SO dump}}
\end{equation}
\begin{equation}
\textrm{Relevance}~\nu(t) = \frac{\# \textrm{of Questions with tag $t$ in $\mathcal{P}$}}{\# \textrm{of Questions in $\mathcal{P}$}}
\end{equation}
Here, $\mathcal{P}$ denotes all questions with tags $\mathcal{T}_{init}$.
Our 49 experiments with a 
broad range of threshold values of $\mu$  and $\nu$ 
show that $\mu = 0.3$ and $\nu = 0.001$ allow for a more 
relevant set of IoT tags. The threshold values  are consistent with 
previous work~\cite{yang2016security,Ahmed-ConcurrencyTopic-ESEM2018}. We finally got 75 IoT tags, including security-related tags. \it{\ul{Third}}, we found a total of 81,651 posts (questions and answers). Out of which, around 48\% are questions (i.e., 39,305) and 52\% (42,346) are answers. Out of the answers, around 33\% (13,868) are accepted. Following previous
research~\cite{Bagherzadeh-BigdataTopic-FSE2019,Barua-StackoverflowTopics-ESE2012},
we only consider the questions and accepted answers to the questions to avoid noise.
Our final dataset $\mathcal{B}$ consists of a total of 53,173 posts (39,305 questions, 13,868 accepted answers).

Given our focus is to study IoT security discussions, it is important that we can pick as many IoT related discussions as possible that could be relevant to IoT security. Intuitively, if an IoT developer is worried about security of IoT, he/she may add `security' or `iot security' as a tag in SO. However, we did not find any tag as `iot security' in SO. We found a tag `security' in IoT related posts in SO, which was accompanied by one of the 75 tags. For example, \dq{48607181} discusses about the possibility of REST API in IoT over SSH. The question has both `iot
 and `security' as tag. However, as we discuss in \sec\ref{sec:results}, most of the IoT posts in SO do not have a `security' tag associated, even if the question contains security-related discussions.

Developer discussion mostly consists of textual discussion, such as how to connect Arduino wifi modules. However, developers sometimes seek programming support and resource sharing links. Sometimes, developers post building logs with images. Thus, a post may contain code-snippets, text, urls, etc. This makes SO posts more complex. In our investigation, we focus on only textual discussion, and thus we remove all types of discussion, such as source codes, logging, urls, etc., except text. We use `beautifulsoup' library to filter out text, source codes, logging sentences, and urls. This step ensures that our dataset contains only text. However, there is a small amount of inline source code that can't be filtered out. For example, the SO question, \dq{7383606}, has the following line: \textit{``...It was suggested that I try char msg[] = myString.getChars();...''}. After nltk tokenization, we find that only a few sentences contain such inline codes. Therefore, our IoT dataset is only applicable for textual discussion.

\subsection{Creation of IoT Security Aspect Benchmark}\label{iot-dataset-creation}

\begin{table}[t]
  
  \caption{Summary stats and agreements for the IoT Security dataset}
    \begin{tabular}{r|ll|rr}\toprule
    \multicolumn{3}{c}{\textbf{Summary}} & \multicolumn{2}{c}{\textbf{Agreement}} \\
    \cmidrule{1-5}
          \textbf{Dataset} & \textbf{Sentences } & \textbf{\#Security } & \textbf{Kappa} & \textbf{Percent} \\
          \midrule
    Benchmark & 5919   &   1049 (17.7\%)     & 0.94     &  97.6\%  \\
    Judgemental & 384 & 40 (10.5\%) & 0.88 & 95.6\% \\
    
    
    

    \bottomrule
    \end{tabular}%
  
  \label{tab:validation-sample-stat}%
\end{table}%
We study the performance accuracy of the studied DL transformer models to detect IoT security-related sentences in SO. Thus, we create a benchmark dataset of 5,919 sentences from our IoT dataset of 53K posts. 
First, we use nltk~\cite{website:nltk} to segregate sentences from the 53K IoT posts, which 
results in 672,678 sentences (i.e., an average of 13 sentences per post). Then we randomly select 5,919 non-overlapping sentences and label those.
This sample size is statistically significant with a 99.99\% confidence level and
2.5 confidence interval. The first two authors then manually label each sentence as 1 (i.e., contains security discussion) 
or 0 (otherwise). Prior to labeling, the authors consult together to establish a coding guideline. The second author has a PhD and has published peer-reviewed high-impact papers in software security. Following that, each author separately labels all sentences based on the guidelines. The agreement analysis is computed between the authors based on two metrics (Cohen Kappa and Percent Agreement). 
The disagreements are resolved by mutual discussion between the authors. 
Finally, the random samples have 1049 sentences labeled as security (17.7\%). As a post consists of multiple sentences (i.e., security and non-security related sentences) and security-related sentences can be anywhere inside a post, the positive rate (\% of security-related samples) is low. One prior study on SO security was conducted by Uddin et al. \cite{Uddin-OpinionValue-TSE2018}, where there were only 4\% (197 samples) security-related samples out of 4522 samples. In \tbl\ref{tab:validation-sample-stat}, we show summary statistics of the IoT Security dataset. The last two columns of \tbl\ref{tab:validation-sample-stat} show the agreement 
level between the two coders during their manual labeling. The Cohen kappa value is 0.94 and the percent agreement is 97.6\%. According to Viera at el.~\cite{Viera-KappaInterpretation-FamilyMed2005}, a kappa value between 0.81 and 0.99 denotes a perfect agreement.

\subsection{Creation of IoT Judgmental Dataset}
In the previous section, we discussed how our benchmark dataset is created. We train our model on the benchmark dataset and select the best performing model. However, model performance on the benchmark dataset does not give us the guarantee that the model will perform similarly when tested in different environments. Therefore, we generate a new dataset to cross-check the best performing model. To select the size of the dataset, we follow Burmeister et al. \cite{samplesize} and find that a minimum of 384 samples are needed to represent the whole IoT dataset of 672678 samples with a confidence of 95\% and an interval of 5. We first exclude all 5919 samples from the IoT dataset and then collect 384 samples randomly. Finally, we label those 384 samples. We call this newly created dataset as ``Judgemental dataset''. The statistics for this dataset are shown in the Table \ref{tab:validation-sample-stat}. The Judgemental dataset contains approximately 10\% security-related samples. We use this dataset to judge our best performing deep model on the benchmark dataset.  

\subsection{Baseline Models}
In our experiments, we use two baseline models, i.e., Support Vector Machine (SVM) \cite{svm} and Logistic Regression (Logit) \cite{logits} that are used by Uddin et al. \cite{Uddin-OpinionValue-TSE2018} for software aspect classification in SO. For classification tasks, both SVM and Logit provide probabilistic values that lie between 0 and 1 for a given input. However, these models take input as numerical values. As we are dealing with textual content, we first convert text samples into numerical values. We use Term Frequency-Inverse Document Frequency (TF-IDF) \cite{tf-idf} to get a numerical value for a sample text. The inverse proportion of the frequency of a word in a given document to the percentage of documents the word appears in is used by TF-IDF to generate values for each word in a document. Words with high TF-IDF numbers imply a strong relationship with the document they appear in. We convert all samples to numerical values using TF-IDF and then feed this input to the baseline models. All hyparparameters are used similar to Uddin et al. \cite{Uddin-OpinionValue-TSE2018}.
\subsection{Hypertuning of Studied Deep Learning Transformer Models}\label{sec:studied-models}
\begin{table}[t]
  \centering
  \caption{Studied Deep Learning Models}
    \begin{tabular}{lp{11cm}}\toprule
    \textbf{Name} & \multicolumn{1}{l}{\textbf{Rationale to Study}} \\
    \midrule
    BERT  &  BERT is designed to learn pre-trained contextual word representation by running of large volume of unlabeled texts. 
    The transformer mechanism allows the encoder and decoder to see the entire input sequence at once.
    \\
         XLNet &  XLNet offers better training methodology and uses more data for pre-training than BERT, which could increase its accuracy in learning.
          \\
          RoBERTa & RoBERT uses more data and computation power to learn from the data. It also offers better training methodology than BERT and XLNet. As a result, RoBERTa outperforms both BERT and XLNet on benchmark datasets (e.g., GLUE).
          \\
          DistilBERT & DistilBERT is also a lightweight version of BERT in terms of the learned architecture and word embedding, while retaining more than 95\% of BERT performance.
          \\
          BERTOverflow & BERTOverflow is an in-domain pretrained BERT. It is pretrained on SO dumps that include code snippets. As a result, it offers better performance on software domain-specific tasks.\\
          \bottomrule
    \end{tabular}%
  \label{tab:studied-models}%
\end{table}%

\begin{table}
\centering
\caption{Architecture details of variants of BERT Model (L = Layers, H = Hidden, D = Heads, P = Parameters)}
\begin{tabular}{l|p{3.2cm}rrrr}
\toprule
\textbf{Architecture}& \bf{Used Model} & \bf{L}  & \bf{H} & \bf{D} & \bf{P}\\ 
\midrule
 BERT & bert-base-uncased & 12& 768&12&110M\\
 XLNet & xlnet-base-cased & 12& 768& 12& 110M                  \\ 
 RoBERTa& roberta-base & 12&768&12&125M                               \\ 
 DistilBERT &distilbert-base-uncased&  6& 768 & 12 & 66M                                \\
  BERTOverflow& jeniya/bertoverflow&  12&	768&12 &	110M\\
\bottomrule
\end{tabular}
\label{architecture_details} 
\end{table}

We investigate five state-of-the-art transformer models: \begin{inparaenum}[(1)]
\item BERT,
\item RoBERTa, 
\item XLNet, 
\item DistillBERT,
\item BERTOverflow
\end{inparaenum}

Transformers \cite{transformers-kant} are attention-based DL models that can handle sequential data. These trained models are used for various language-specific downstream tasks after fine-tuning the pre-trained models for the specific task-related domain (e.g., sentiment detection, machine translation). Our selection of these five transformer models is based on previous findings in SE classification tasks. BERT has shown massive improvement over other deep models for sentiment analysis tools in SE~\cite{bert-sentiment-se}\cite{bert-se-sentiment}. Other variants of BERT, i.e., RoBERTa, XLNet, etc., are also used in the SE classification task. RoBERTa, the most optimized BERT model, has shown better performance in multiple SE classification tasks. For example, Uddin et al. \cite{uddin-sentiment} found that RoBERTa performed better than other transformer models in six different datasets for sentiment detection. In another recent study, Dai et al. \cite{Dai-roberta} showed RoBERTa's superiority in aspect level software sentiment analysis. Moreover, we find that other transformer models, such as DistilBERT and XLNet, also perform as well as the BERT model \cite{empirical-transformers}. Besides this, Tabassum et al. \cite{tabassum2020code} found that the SO domain specific model BERTOverflow had improved the performance of SO related classification tasks. This motivated us to study these five models for our IoT security aspect classification task. We leave the exploration of recent state-of-the-art models like GPT2 and T5 as our future work. In Table \ref{tab:studied-models}, we show the details of the five studied models in this paper.

\begin{figure}
    \centering
    \includegraphics[scale=.6]{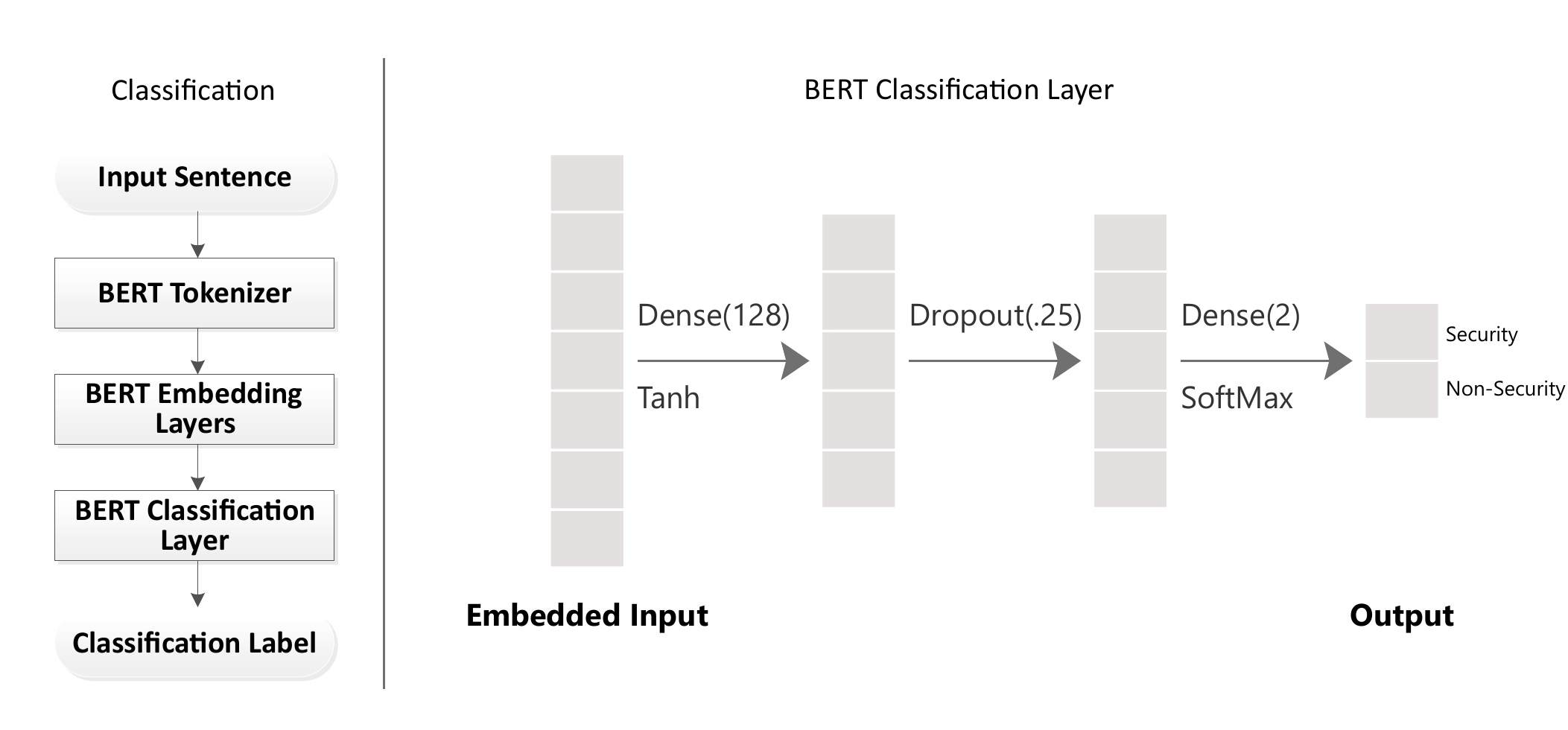}
    \caption{BERT Classification Processing}
    \label{fig:BERT-classifier}
\end{figure}

\begin{table}[t]
\centering
\caption{Hyperparameters of the deep learning models on the IoT Security dataset}
\begin{tabular}{l|rrrrr}
\toprule
\textbf{Model}& \bf{Batch Size} & \bf{Epochs}  & \bf{Optimizer}  & \bf{Learning Rate}\\ 
\midrule
BERT & 32 & 3 & Adam  & 2e-5 \\
RoBERTa & 16 & 2 & Adam  & 1e-5 \\
XLNet & 8 & 3 & Adam  & 2e-5 \\
DistilBERT & 8 & 3 & Adam  & 2e-5\\
BERTOverflow & 32 & 2 & Adam  & 1e-5\\
\bottomrule
\end{tabular}
\label{dllParamAPIReview} 
\end{table}

\nd\bf{Fine-Tuned Architecture.} Table \ref{architecture_details} shows the architectural details of the used Transformer models from the
\href{https://huggingface.co/transformers}{\textcolor{blue}{hugging face}}
transformer library, an open-source community centered on pretrained
transformer libraries. BERT architecture provides the contextual embedding for
each word in a sentence. For classification purposes, a classifier is used
at the top of the model that takes the embedding of sentences as input and
outputs the predicted weights for each category (Security and Non-Security). \fig \ref{fig:BERT-classifier} (right) shows the used architecture of the classifier. It consists of a dense layer with $Tanh$ activation followed by a dropout layer, a dense layer with $SoftMax$ activation. \fig\ref{fig:BERT-classifier} (left) 
shows how we used the BERT architecture for our security aspect detection.
First, a tokenizer takes the input sentences and represents them as tokenized
sentences. Then, these tokenized sentences are fed into the BERT model. Finally, the
embedded sentences are passed through the classifier to get the final
output (security or non-security).

Following the sequence classifier model from \href{https://huggingface.co/transformers}{\textcolor{blue}{hugging face}}, we apply $Tanh$ activation. In addition, we use the dropout layer to reduce overfitting. To finetune the model on our dataset, we initialize the classifier randomly where the BERT is initialized from pretrained models.

For each deep learning model, following literature in text classifications~\cite{Alharbi-LSTM-CognitiveJournal2019,Hameed-BiLSTM-IEEEAccess2020,Delvin-BERTArch-Arxiv2018}, 
we tuned the following hyperparameters for optimal 
performance: \begin{inparaenum}[(1)]
\item Batch size, 
\item Number of hidden layers, neurons, units, filters, epochs 
\item Optimizer, Activation, and loss functions, and
\item Learning rate.
\end{inparaenum} To tune these hyparparameters, we use standard Grid search technique~\cite{random-search}. We select a set of batch sizes (i.e., 8, 16, and 32), epochs (i.e., 2, and 3), and learning rates (i.e., 1e-5, 2e-5, and 3e-5) as suggested in \href{https://huggingface.co/}{\color{blue}hugging face}. We run the models on the dataset for each set of hyperparameters. Then we choose the best hyperparameter set for each model based on their performances. For example, we find the best performance of BERT for batch size of 32, epoch of 3, and learning rate of 2e-5.
In \tbl\ref{dllParamAPIReview}, we show the values of selected hyperparameters in our deep learning model after tuning.  
Our replication package details the hyperparameter values with source code.


%

\nd\bf{{Data Processing.}} We prepare our dataset with only textual discussion. However, there may be some urls. These unwanted texts may hinder the model's performance. Thus, we check for URLs and remove them using regular expressions. For implementation, we use Python'regex' library. In addition to this, we remove all stop words. As the importance of stop words is low and the frequency of these words is high, these words can obscure other significant words. Moreover, recent studies related to sentiment analysis found that stop words removal is an effective step in improving a model's performance \cite{stop-wprds}. Therefore, we remove all stop words using `nltk' \cite{website:nltk} library.

\nd\bf{{Model Features.}} We used the pre-trained embedding in BERT models, which learns 
additional contextual information for a given domain (e.g., in our case, security in developer discussions) during the training phase. The combination of pre-trained and domain-specific learning into such pre-trained models offers more contextual information during classification tasks than traditional DL models. We add `EOS' and `CLS' to each sentence and then tokenize all sentences using a pretrained BERT-Tokenizer. We
padded each sentences to 100 length with `zero' padding.

%
%


\subsection{Performance Metrics} \label{sec:performance-metrics}
We analyze the performance of the classification models for security aspect detection using five standard metrics in 
information retrieval~\cite{Manning-IRIntroBook-Cambridge2009}: Precision (P), Recall (R), F1-score (F1), Accuracy (Acc), and Matthews Correlation Coefficient (MCC). 
\begin{enumerate}[leftmargin=20pt]
  \item Precision (P) is the fraction of sentences labeled as `HasSecurity' (i.e., the sentence contains security discussions) out of all the sentences labeled as `HasSecurity'. 
  \begin{equation}
  P  = \frac{TP}{TP+FP}
  \end{equation}
Here TP = Correctly classified a sentence as containing discussion about security, FP = Incorrectly classified as containing discussion about security,   
  \item Recall (R) is the fraction of all `HasSecurity' sentences that are successfully identified out of all sentences that contain security discussions.
  \begin{equation}
	R = \frac{TP}{TP+FN}
  \end{equation}
  Here FN = Incorrectly classified as not containing discussion about security.
  \item F1-score is the harmonic mean of precision (P) and Recall (R). 
  \begin{equation}
	F1 = 2*\frac{P*R}{P+R}
  \end{equation}
  \item Accuracy (Acc) is the fraction of all correctly classified out of all records.
  \begin{equation}
	Acc = \frac{TP + TN}{TP + TN + FP + FN}
  \end{equation}
  Here TN = Correctly classified as not containing discussion about security.
  \item Matthews Correlation Coefficient (MCC) is the linear correlation between all `HasSecurity' labeled sentences and the sentences that contain security discussion. It takes into account all possible scenarios- i.e., correctly identified security-related sentences, incorrectly identified security-related sentences, correctly identified non-security-related sentences, and incorrectly identified non-security-related sentences.
  \begin{equation}
	MCC = \frac{TP * TN - FP * FN}{\sqrt{(TP + FP)(TP + FN)(TN + FP)(TN + FN)}}
  \end{equation}
  \item Area Under the Curve (AUC) is the area under the Receiver Operating Characteristics (ROC) curve. An ROC curve is a graphical plot to illustrate the ability of a binary classifier (such as ours). 
  $$ TPR = \frac{TP}{TP + FN}, ~
  FPR = \frac{FP}{FN + TP}$$
  \begin{eqnarray}
  AUC = \int_{x = 0}^{1} {TPR}({FPR}^{-1}(x))dx
  \end{eqnarray}
  
\end{enumerate}
F1-score is preferred over metrics like accuracy (Acc) when evaluation dataset is imbalanced. For example, 
our IoT benchmark dataset is imbalanced with lower number of sentences labeled as 1 (i.e., `HasSecurity' = 1) than sentences labeled as 0.  
Following standard practices in Literature~\cite{Zhang-SentimentDetectionSEBERT-ICSME2020,Uddin-OpinionValue-TSE2019}, 
we use F1-score to determine and report the best model.

\section{Empirical Study Results}\label{sec:results}
In this section, we answer four research questions (RQ) using SO data dump of September 2019. We compare the results against those found in SO data dump of January 2022 in \sec\ref{sec:discussion}. 
\begin{enumerate}[leftmargin=50pt, label=\bf{RQ\arabic{*}.}]
    \item How  do  the  deep  learning pre-trained transformer models  perform  to  detect  IoT security aspects in the benchmark? (\sec\ref{sec:rq1})
    \item What  are  the  error  categories  in  the  misclassified  cases  of the best performing model? (\sec\ref{sec:rq2})
    \item What are topics in the IoT security discussions collected from the entire IoT SO dataset using the best performing model? (\sec\ref{sec:rq4})
    \item How do the IoT security topics evolve in SO? (\sec\ref{sec:rq5})
\end{enumerate}

\subsection{RQ$_1$ How  do  the  deep  learning pre-trained transformer models  perform  to  detect  IoT security aspects in the benchmark?}\label{sec:rq1}
\subsubsection{Motivation} 
 Pretrained transformer models have already shown dominating performances over shallow and deep models in software engineering \cite{Zhang-SentimentDetectionSEBERT-ICSME2020}. Moreover, recent studies on software engineering have found the effectiveness of domain-specific embedding (e.g., BERT, RoBERTa, XLNet, etc.) than a generic one. For example, Tabassum et al. \cite{tabassum2020code} shows that the BERTOverflow model that is pre-trained on Stack Overflow dumps beats the performance of efficient transformers like RoBERTa on developer discussions such as Github and Stack Overflow. Therefore, we investigate the possibility of improving the baseline performance to detect security aspects in IoT discussions using studied Transformer Models, e.g., BERT. 

\subsubsection{Approach} \label{sec: rq1-approch}
\begin{figure}
    \centering
    \includegraphics[scale=.42]{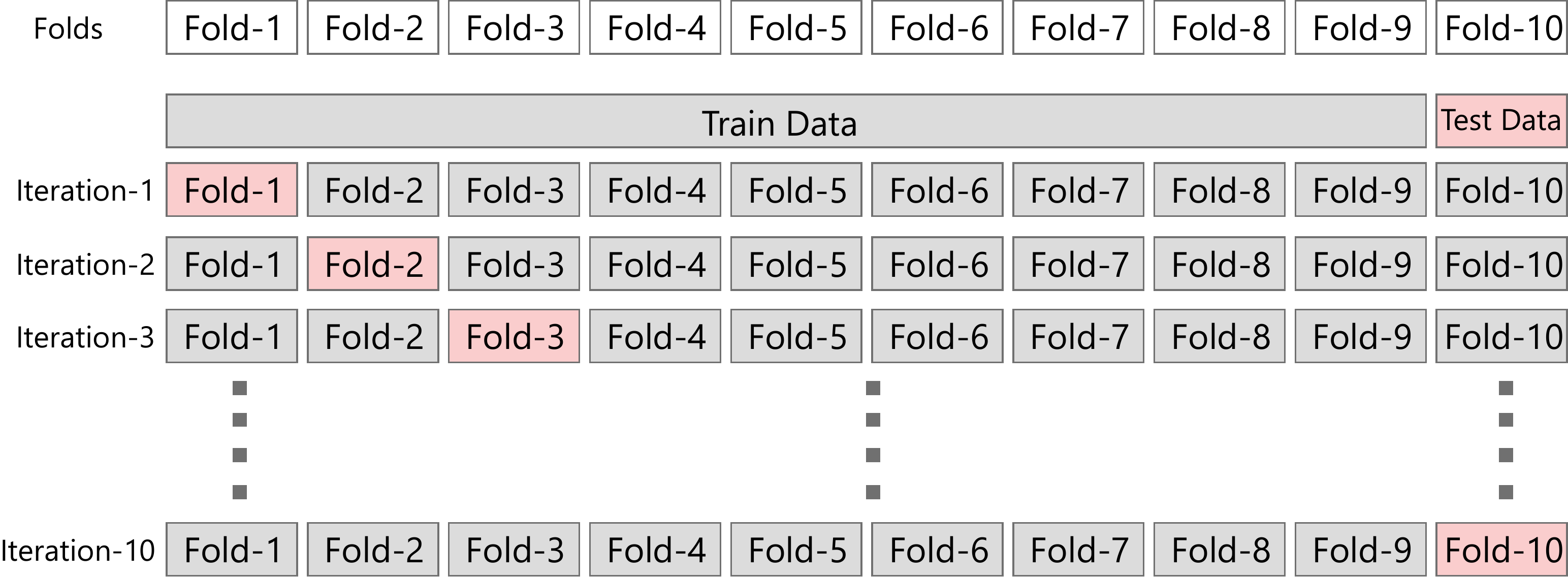}
    \caption{10-Fold cross validation technique used in our study}
    \label{fig:k-fold}
\end{figure}
We first use the newly created IoT Security dataset 
from \sec\ref{iot-dataset-creation} to get the best performing transformer model. For experimentation, similar to Uddin and Khomh~\cite{Uddin-OpinionValue-TSE2019}, we use a stratified k-fold cross-validation ( k = 10). Each fold divides dataset $D$ into $k$ disjoint folds, each of
approximately equal size.
$D_i\cap D_j=\phi~~~\forall~i,j\in k~and ~i\ne j $ and 
$D_1\cup D_2\cup...\cup D_k = D$. To avoid overfitting: we train the models $k$ times each time we hold one of the $k$ folds for validation, and we use the rest $k-1$ folds for training. The value of $k$ is a hyperparameter. In our experiments, we set K as 10. In \fig \ref{fig:k-fold}, we show 10-Fold cross validation techniques. We run the model 10 times. In each iteration, we train the model on nine different folds and test the trained model on the last fold. Then, we calculate the average performance of these 10 iterations. For example, according to the \fig \ref{fig:k-fold}, at the first iteration, we pick `Fold-2', `Fold-3', `Fold-4', `Fold-5', `Fold-6', `Fold-7', `Fold-8', `Fold-9', and `Fold-10' to train the model, and we test this model on the `Fold-1'. Therefore, in our case, we fine tune our model during each training (i.e., k times) and then test the tuned model on the test fold. 
To tune the hyperparameter, we follow grid search \cite{Shekar-grid-search} tuning techniques. For example, we set learning rate as 1e-5, 2e-5, and 3e-5 and select the one that has highest performance for each model. For the two baselines (Logit and SVM), we hyper tuned all the parameters used by Uddin and Khomh~\cite{Uddin-OpinionValue-TSE2019,Uddin-OpinerReviewAlgo-ASE2017}. We find the performance of transformer models on benchmark dataset and select the best performing model based on the F1-Score. Then, we test the trained model on our Judgmental dataset to check either the model learn to generalize the security discussion in IoT domain.

\begin{table}[t]
  \centering
  \caption{Performance of security aspect detection of deep transformer models in the IoT Security dataset~\cite{Uddin-OpinionValue-TSE2019}}
  \resizebox{1\columnwidth}{!}{%
    \begin{tabular}{p{20mm}lrrrrrr}\toprule
    \textbf{Type} & \textbf{Model Name} & \textbf{Acc} & \textbf{P} & {\textbf{R}} & {\textbf{F1}} & {\textbf{AUC}} & {\textbf{MCC}} \\
    \midrule

    Transformer & BERT  & 0.976 & 0.928 & 0.939 & 0.934 & 0.962 & 0.919 \\
          & RoBERTa & \bf{0.977} & 0.930 & \bf{0.941} & \bf{0.935} & \bf{0.962} & \bf{0.922} \\
          & XLNet & 0.972 & 0.922 & 0.921 & 0.921 & 0.952 & 0.904 \\
          & DistilBERT & 0.969 & 0.907 & 0.921 & 0.914 & 0.950 & 0.895 \\    
    & BERTOverflow & 0.976 & \bf{0.938} & 0.928 & 0.933 & 0.957 & 0.919 \\
    \midrule
    Baseline & SVM   & 0.953 & 0.876 & 0.856 & 0.866 & 0.915 & 0.838 \\
                     & Logit & 0.952 & 0.875 & 0.853 & 0.864 & 0.913 & 0.835 \\          \midrule
    \multicolumn{7}{c}{\bf{Performance Change in RoBERTa Over Baselines}} \\
    \midrule
    \multicolumn{2}{c}{Over SVM} & 2.5\% & 6.2\% & 9.9\% & 8.0\% & 5.1\% & 10.0\%\\
    \multicolumn{2}{c}{Over Logit} & 2.6\% & 6.3\% & 10.3\% & 8.2\% & 5.4\% &10.4\%\\
    \bottomrule
    \end{tabular}%
    }
  \label{tab:modelPerformanceBERTAIoTSecurityDataset}
\end{table}%



\subsubsection{Results} 
In \tbl\ref{tab:modelPerformanceBERTAIoTSecurityDataset}, we report the performance of two baselines and the five advanced language-based pre-trained transformer models on the IoT security dataset. For each model,
we report six performance metrics from \sec \ref{sec:performance-metrics}.
Among the two shallow learning baselines, SVM offers the best performance for all six metrics, outperforming Logits. SVM shows an F1-score of 0.866, while Logits shows an F1-score
of 0.864. For the other five performance metrics - Accuracy, Precision, Recall, AUC, and MCC, the improvements are around .01\%, .01\%, .02\%, .02\%, and .04\% respectively. We note that the performance for both models is high. The higher performance of these models on our dataset is due to the following reasons: Classifications can often be more straightforward for security aspects, i.e., parsing of words could suffice (e.g., secure, attack, and hacking). So, non-contextual models like SVM and Logits can detect security aspects with high precision and recall. 

All five transformer models show better performances in terms of all metrics than the baseline models from Uddin \cite{Uddin-OpinerReviewAlgo-ASE2017} (SVM and Logit). The last two rows in \tbl\ref{tab:modelPerformanceBERTAIoTSecurityDataset} show the performance change (in percentages) we observed in the best performing transformer model (RoBERTa) compared to the two baselines. For example, the best 
performing transformer model is RoBERTa with an F1-score of 0.935, which outperforms the baselines SVM by 8.0\%, and Logits by 8.2\%. Higher performances mean that transformers are precisely (i.e., around 6\% higher precision) discovering (i.e., around 10\% higher recall) security-specific sentences. It is because these models can understand the contextual meaning in a sentence. For example, the following non-security related sentence \emt{My goal is to create a sort of password-like system where if you push the right buttons in the right order then a button will light up.} The comments describe the procedure of a system using the references `password like' that have no relation with security. The baselines model erroneously labeled it as 0 due to the misleading keyword- `password like' where the transformer models discerned the context. Hence, it found no security threats in this sentence. This superiority of these models over baselines is not surprising given that the transformer models are generally found to have offered better performance than state-of-the-art classifiers for multiple other text classification tasks in software engineering (e.g., sentiment detection~\cite{Zhang-SentimentDetectionSEBERT-ICSME2020}).

Among the transformers, RoBERTa shows the best performance with an F1-Score of 0.935, while DistilBERT is the lowest with an F1-Score of 0.914. F1-Score for the other three models- e.g., BERT, BERTOverflow, and XLNet are 0.934, 0.933, and 0.921 respectively. There is a tiny gap between the performance of RoBERTa and BERT. It is because the architecture and training procedures of these two models are almost similar. The modification, dynamic changes of the masking pattern (in RoBERTa) instead of static masking (in BERT) during the training phase, could be the reason behind the subtle improvement in performance. 
XLNet improves the performances by increasing the training to handle the dependency between the words and capture long-distance information. In our dataset, each sentence is simple and small in size as well. So, large-scale training on the dataset causes overfitting and lower performance than BERT. The performance drop of DistilBERT is understandable as the distillation of the BERT model is obviously lower performing than its original model.

Although BERTOverflow has the same architecture as BERT and is trained on the StackOverflow dumps, the performance of this model is lower than RoBERTa and BERT. RoBERTa and BERT model beat BERTOverflow by 0.21\% and 0.11\% in terms of F1-Score. However, BERTOverflow has the best precision of 0.938 among the transformer models. This result indicates that BERTOverflow predicts security-specific sentences precisely (i.e., better precision) but lacks in dicoverability (i.e., lower recall and F1-Score) in comparison to RoBERTa and BERT. We look into the cause of BERTOverflow's performance drop. Our finding in this scenario is that BERTOverflow pretrained on StackOverflow dumps containing
code snippets, so the model faced some security-related words (e.g., 'private`,
'protected`, 'bug`, 'break`, 'lock`, 'key`) in more programming contexts than
security contexts. The model fails to learn the correct embedding for these
words during the fine-tuning or training phase. For example, consider the
following security-specific sentence \textit{``It will be helpful if anyone states which
are keys to lock so that in addition in my python script for locking phone
programmatically.”} BERTOverflow failed to embed security context for this
sentence and labeled it as 0- i.e., non-security.

\begin{table}
\caption{Spearman Rank Correlation Coefficient for RoBERTa and BERT pair and RoBERTa and BERTOverflow pair.}\label{correlation}
\centering
\begin{tabular}{ccc}
\hline
\textbf{Model Pair} &  \textbf{Correlation Coefficient, $\rho$} & \textbf{p-value}\\
\hline
RoBERTa \& BERT & 0.941 & $ <0.0001$ \\
RoBERTa \& BERTOverflow & 0.940 & $<0.0001$\\
\hline
\end{tabular}
\end{table}

Our intuitions indicate that RoBERTa is the best security-discussion detector, which is also supported by the experimental results. However, RoBERTa has only a 0.1\% and 0.2\% lead over the second and third best models, which is not a significant difference. Thus, we perform a statistical analysis to understand the difference in results between RoBERTa and the other two models. We calculate the Spearman Rank Correlation Coefficient \cite{spearman} for the results of RoBERTa and BERT, and RoBERTa and BERTOverflow. We show the result in Table \ref{correlation}. For both pairs, the correlation scores are as high as 0.94. However, the results of RoBERTa and BERT are more correlated (0.1\% higher) than the results of RoBERTa and BERTOverflow. Following Leclezio et al. \cite{spearman2}, the scores indicate that the results of RoBERTa are strongly correlated with the results of BERT and BERTOverflow. Nevertheless, the score also indicates that there is a difference between the results of RoBERTa and other models, as the score is far less than 1.

\begin{figure}[t]
  \centering
  \includegraphics[scale=.5]{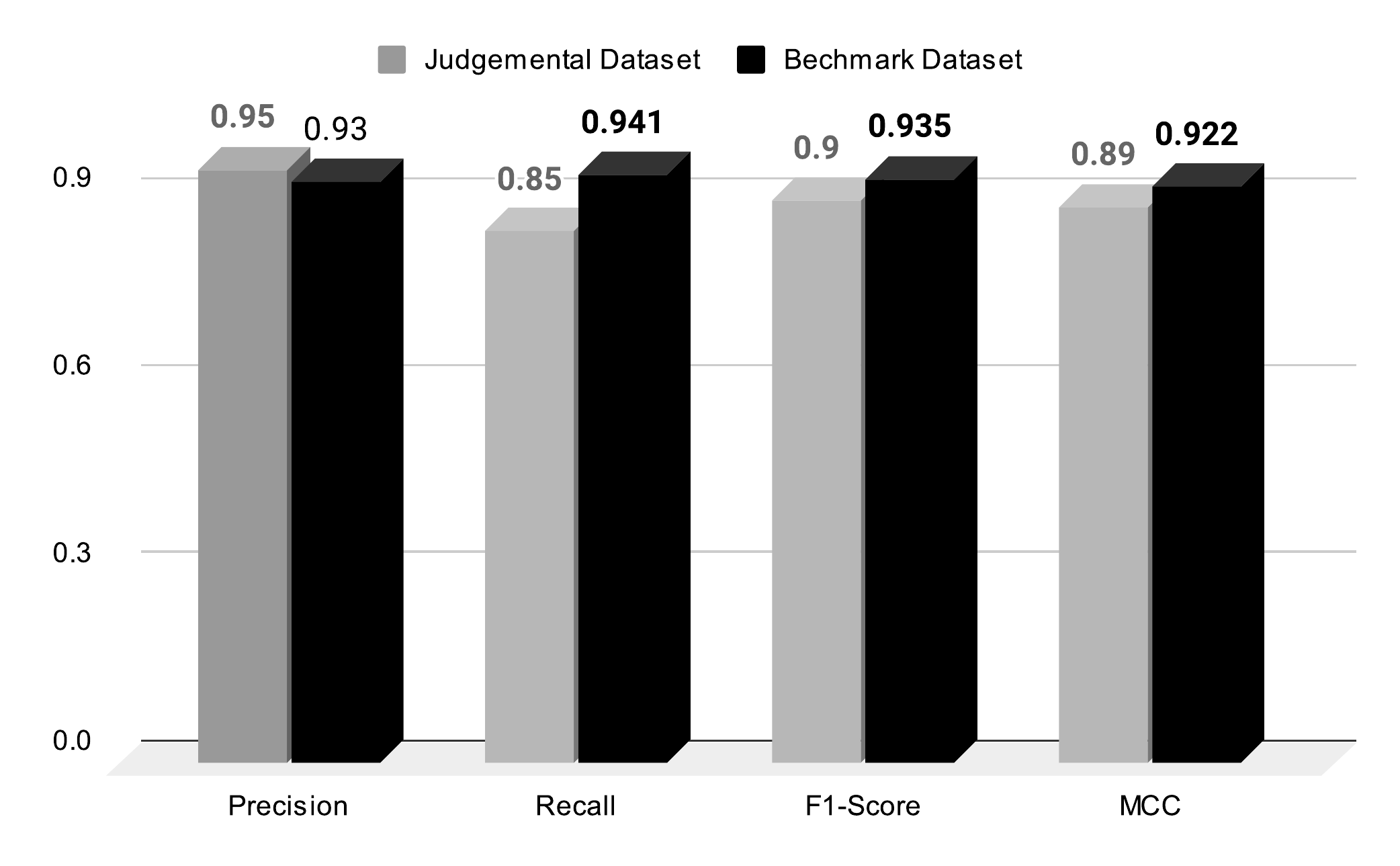}

  \caption{Performance of RoBERTa on bechmark and judgemental dataset}

  \label{fig:compare-judgemental-benchmark}
\end{figure}

From these experiments and analysis of the results, we find RoBERTa is the best performing deep learning model. We further test this trained RoBERTa model on our judgemental dataset. We find that the model performs similarly on the judgemental dataset. RoBERTa has a precision of 0.95, a recall of 0.85, and a F1-Score of 0.9. We report the performance in the \fig \ref{fig:compare-judgemental-benchmark}. Compared to the performance on the benchmark dataset, model has better precision but lower recall, which results in a slightly lower F1-Score on the judgemental dataset. This result indicates that the model has learned to generalize security discussion in IoT domain. Thus, our RoBERTa model is reliable enough to carry on further experiments. From now on, we refer to this best-performing transformer model, RoBERTa, as \bf{\ul{SecBot}} in the remainder of the paper.

\subsection{RQ$_2$ What  are  the  error  categories  in  the  misclassified  cases  of the best performing model?} \label{sec:rq2}

\subsubsection{Motivation}
Our observation in the previous research question (RQ$_1$) is that the most robust generic model Secbot (RoBERTa) outperforms domain-specific BERTOverflow. Although improvement is subtle, the statistical significance in Table \ref{correlation} suggests there are enough differences in both model's predictions. Our intuition behind this RQ is that lack of domain knowledge in Secbot (RoBERTa) should have some impact on its performance. Thus, we perform a qualitative study of the misclassifications of the best-performing model (i.e., SecBot) to detect security aspects in IoT discussions. Proper reasoning of misclassification will strengthen our claim that the model performance is not arbitrary. Moreover, this gives us leverage to confine our model's vulnerabilities in detecting security aspects in IoT domains. In the future, this will provide a scope for the researcher to learn about the possible avenues for improvement of the model.
\subsubsection{Approach}\label{rq2-approach}
In our benchmark of 5,919 samples, SecBot was wrong in 139 cases (false positive or false negative). We analyzed these 139 samples from our IoT security dataset as follows. We applied 10-fold cross-validation to our entire 
dataset following steps similar to \sec\ref{sec: rq1-approch}. 
As such, we were able to use the entire 5,919 sentences for training and testing. At the end of the $10^{th}$ iterations, we find a total of 139 misclassifications in the IoT dataset by the RoBERTa model used for SecBot. Then we manually analyze all the 139 sentences to determine the reason for misclassification. The manual analysis process consists of three stages. Both of the authors consulted together (over Skype) on all 139 sentences to give a textual label that could explain the reason for misclassification. For example, the following sentence is erroneously labeled as 1 (i.e., contains security discussion) by SecBot \emt{I have attempted to set it so that it loads as the root user each time with no luck.} 
After careful comparison of this sentence with other sentences in our IoT samples, we determine that the misclassification is due to the ambiguity in context, which arose due to the words- `attempted to set' and `root user' in the sentence. We label the misclassification reason as `Ambiguous Context'; 
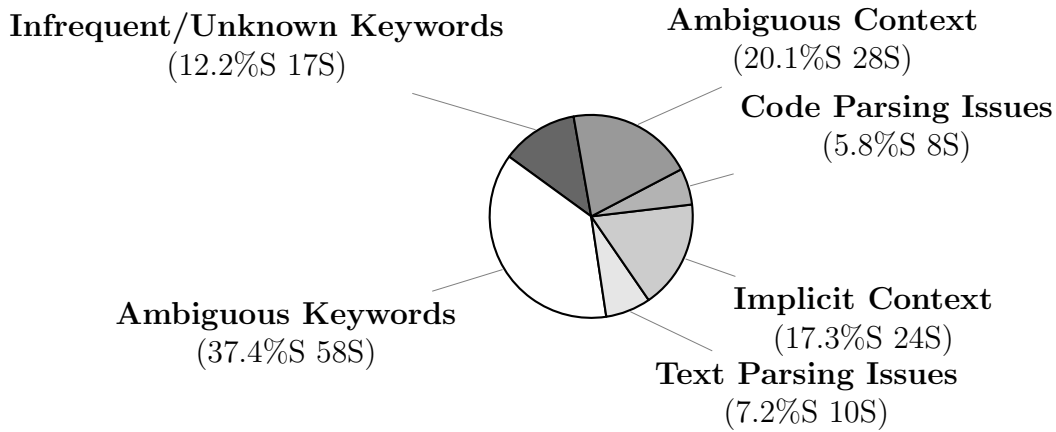
\begin{figure}[t]
	\centering\begin{tikzpicture}[scale=0.45]-
    \pie[
        /tikz/every pin/.style={align=center},
        text=pin, number in legend,
        explode=0.0, rotate=100,
        color={black!60, black!0, black!10, black!20, black!30, black!40},
        ]
        {
            12.2/\bf{Infrequent/Unknown Keywords}\\(12.2\%S 17S),
            37.4/\bf{Ambiguous Keywords}\\(37.4\%S 58S),
            7.2/\bf{Text Parsing Issues}\\(7.2\%S 10S),
            17.3/\bf{Implicit Context}\\(17.3\%S 24S),
            5.8/\bf{Code Parsing Issues}\\(5.8\%S 8S),
            20.1/\bf{Ambiguous Context}\\(20.1\%S 28S)
        }
    \end{tikzpicture}
	\caption{Distribution of SecBot error categories in the total 139 sentences (S) from IoT Security dataset}
	\vspace{-2mm}
	\label{fig:secbot-error-category-total}
\end{figure}
\subsubsection{Results}




In \fig\ref{fig:secbot-error-category-total}, we show the misclassification categories in all 139 sentences collected from the IoT Security dataset. 
Overall, we find six error categories: \begin{inparaenum}
\item Ambiguous Keywords, 
\item Ambiguous Context, 
\item Implicit Context,
\item Infrequent/Unknown Keywords,
\item Text Parsing Issues, and
\item Code Parsing Issues.
\end{inparaenum} The most observed error category is `Ambiguous Keywords', which we observe in 37.4\% of all 139 sentences (i.e., 58 sentences). 
The category `Ambiguous Context' is observed the second most times (20.1\% sentences), followed by the category `Implicit Context' (17.3\% sentences), `Infrequent/Unknown Keywords' (12.2\% sentences), `Text Parsing Issues' (7.2\% sentences) and 
Code Parsing Issues (5.8\% sentences), respectively. We discuss the error categories below.

\begin{inparaenum}[(1)]
\item \bf{Ambiguous Keywords.} This type of error occurred due to the ambiguity of a word in the dataset. We found two scenarios where this type of error occurred. When a sentence offers insufficient context to detect security concerns, thus, prediction mostly depends on the keywords present in that sentence. In the worst-case scenario, these keywords are found in both security and non-security contexts during training. The model hardly guessed correctly when this intriguing scenario occurred. For example, SecBot erroneously labeled the following sentence as 1, due to the ambiguous presence of keywords like `signed'- \emt{I'm not sure how to handle signed bytes.} It got confused as `signed' word associated with various security-related terms (e.g., `signed device', `self-signed cert', `signed aws') and also non-security specific sentences (e.g. `signed bit', `signed number') as well. So, the model erroneously predicted this programming-related non-security sentence as a security-specific sentence. Secondly, this error occurred when a keyword was predominantly associated with security-specific sentences during training and pretraining phases. But in the erroneous cases, the keyword was found in a non-security specific scenario or vice versa. Consider the sentence- \emt{I am trying to add the option to change the wi-fi password, but that part of the form does not seem to be available to the code. } SecBot labeled this sentence as 1 (i.e., security), but the underlying scenario is about the configuration of wifi. SecBot was confused due to the keywords `password' and `change' --which are predominantly found in the security-specific sentences during training.

\item \bf{Ambiguous Context.} These errors occur when the underlying context is too ambiguous to understand, i.e., security-specific. Such ambiguity occurred because more information from surrounding sentences was needed. Consider the following sentence: \emt{I hope that someone here can help me, I want to control my arduino uno by sending it commands from a c++ program that performs some basic face recognition.} SecBot labeled the sentence as 1, while it is 0 in the manual label. It is because the context is about controlling an IoT device that performs face recognition using some commands. While the manual labeling phase considered the sentences non-security, the model found it a possible security breach due to the context of controlling a security device.

\item \bf{Implicit Context.} SecBot was unable to discern the underlying security context when it was implicit. Consider the following sentence: \emt{The main difference from software development standpoint of ios private api that such apis aren't declared in header files.} Here, the context is about ios private API which can only accessed and used by Apple developers. The elementary context is that developers can not use any API without importing, but implicitly it means that this API is restricted for non-Apple developers. As such, SecBot labels it as 0, whereas the benchmark manual label is 1.

Sentences with implicit security contexts are difficult to capture for a fully automated system as domain knowledge experts can only bisect them from other sentences. 

\item \bf{Infrequent/Unknown Keywords.} This type of error occurred when the model encountered a word that was not present during the training or was present so sporadically that it could not learn the word correctly. But that word had profound significance in understanding the sentence. For example, the following sentence is labeled as 0 by SecBot: \emt{The user should have a iam access.}. Here, `iam access'-- an AWS access token—is the keyword to classify the sentence correctly. SecBot encountered this unknown keyword during the testing phase and misclassified the sentence. Consider another instance: \emt{See the hid usage tables, v1.12 document for more information about hid usage values.} The sentence contains a less frequent word, `hid' is present. SecBot could not learn the weights for this keyword correctly, and so it labeled the sentence as 0, where manual labeling considered it as 1.

\item \bf{Text Parsing Issues.} This error occurred due to the incompetence of the model to parse the textual contents correctly in a sentence. For example, SecBot erroneously labels the following sentence as 0, even though it is about the enablement of secure socket layer (SSL) in the Raspbian Stretch device: \emt{The version of Openssl on Raspbian Stretch is 1.0.2, but according to Openssl website, these versions should be binary compatible (https://www.openssl.org/policies/releasestrat.html).} While SecBot knew from training that SSL is related to the security aspect, SSL in this sentence is denoted as Openssl, which is unknown to SecBot.

\item \bf{Code Parsing Issues.} These errors are very few compared to the other categories. This error occurred due to the presence of code snippets and inline code in the samples. For example, SecBot is unable to determine security threats related to input injection noted in the code block that is provided as text in this sentence: \emt{In Package tag add \ldots In Capabilities add \ldots <rescap:Capability Name="inputInjection" /><rescap:Capability Name="inputInjectionBrokered" />}. This type of parsing error is inevitable for automated models like SecBot because security-specific context can be buried anywhere in a code snippet.
\end{inparaenum}

\subsection{RQ$_3$ What are topics in the IoT security discussions collected from the entire IoT SO dataset using the best performing model?}\label{sec:rq-iot-sec-topic}\label{sec:rq4}
We automatically detect security topics in the sentences labeled as security-specific by SecBot in our IoT dataset.  
\subsubsection{Motivation} 
The rapid growth and usage of IoT devices, tools, and technologies have become prominent threats to data protection, security, and privacy. IoT developers encounter various security challenges while using these devices and tools. It is vital to understand those challenges in order to improve and design more secure devices and tools. So, the understanding of topics in the SO IoT security discussion can be helpful to infer those challenges.
\subsubsection{Approach} \it{\ul{First}}, we apply SecBot on the entire 672,678 sentences that we collected from our 53K
IoT posts (\sec\ref{sec:iot-data-collection}). SecBot returned 30,595 sentences
as 1, i.e., those containing security discussions. \it{\ul{Second}}, we preprocess this dataset as
follows. We removed stopwords, extreme words (e.g., words with less than 20 frequencies), most frequent words (e.g., words with more than 2000 frequencies), short tokens, and letter accents. The stopwords are collected from 
NLTK~\cite{website:nltk}. We applied lemmatization to sentences to get root form 
of each word. \it{\ul{Third}}, we fed the preprocessed sentences into the topic modeling algorithm Latent Dirichlet Allocation (LDA)~\cite{blei2003latent} 
available via Gensim~\cite{Radim-gensim-LREC2010}. LDA takes a
set of texts as input and outputs a list of topics to group the texts into K
topics. This K is a hyperparameter of the algorithm. To get the optimal value of K,
we used standard practice as originally proposed by Arun at
el\cite{Arun-OptimalTopic-AKDD2010}. This approach suggests that the optimal number
of topics has the highest coherence among them, i.e., the more coherent the
topics, the better they can encapsulate the underlying concepts. To determine
this coherence metric, we use the standard c\_v metric as originally 
proposed by R\"{o}der et al.~\cite{Roder-TopicCoherence-WSDM2015}. We use the topic coherence metric as available in 
Gensim package~\cite{Radim-gensim-LREC2010}. To determine our optimal number of topics, we use grid search technique. We fist
select a set of values for K. Then we run our LDA
algorithm for each value of K and measured the coherence metric. Finally, we
choose the value of K that is associated with the highest coherence score. In our
experiment, we set K = {1,4,8,9, 10,11, 12, 16,20,24,28,32, 36, 40} and got maximum coherence for K = 9. Thus, we picked 9 as our optimal number of topics. LDA also uses two hyperparameters, $\alpha$ and $\beta$ to control the 
distribution of words in topics and the distribution of topics in posts following Biggers et al.
~\cite{Briggers-ConfigureLDA-EMSE2014}. 
Following previous work~\cite{Barua-StackoverflowTopics-ESE2012,Bagherzadeh2019,Ahmed-ConcurrencyTopic-ESEM2018,Briggers-ConfigureLDA-EMSE2014,Rosen-MobileDiscussionSO-EMSE2016,yang2016security}, we thus use standard 
values $50/K$ and 0.01 for the two hyper parameters. \it{\ul{Fourth}}, we label each topic a name by manually
analyzing a random sample of at least 50 sentences per topic. We do not limit ourselves to a finite number of sentences per topic during the labeling to ensure that the assigned label is reliable. The first and second authors virtually meet to label the topic manually. First, both authors review a number of sentences for a topic and come up with possible topic labels. Then both authors share their thoughts on the possible topic labels. If both authors have the same thoughts, they label the topic with a suitable title; otherwise, they go over the sentences again and label the topics together. For example, the following sentence is about the certificate validation for different users, \emt{\ldots Found the answer as I was trying to import the certificate with the Root and was testing with a different user.} Another sentence assigned to the same topic is \emt{ \ldots The commands where I input the SSID and Password are the ones causing trouble.}. We thus label the topic as `User/Device Authorization'.
\it{\ul{Fifth}}, we assign each topic a category, where a
category denotes a higher order grouping of the topics. For example, we assign the topics `User/Device Authorization' and `Crypto/Encryption Support' to the category `Software', because 
the topics contain discussion about the support of software in IoT devices to ensure security. The topic `Secure Transaction' is assigned to the category `Network, because 
it contains discussions about the securing of transactions/communications between IoT devices/services over the network.

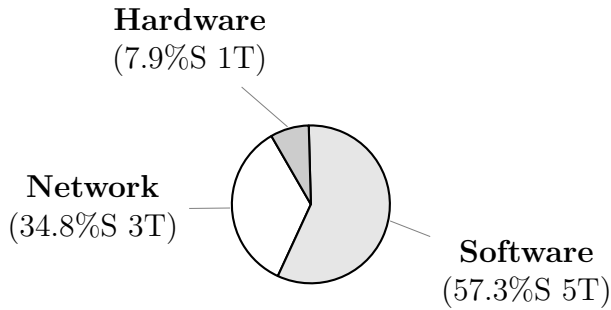
\begin{figure}[t]
	\centering\begin{tikzpicture}[scale=0.35]-
    \pie[
        /tikz/every pin/.style={align=center},
        text=pin, number in legend,
        explode=0.0, rotate=120,
        color={black!0, black!10, black!20},
        ]
        {
            34.8/\bf{Network}\\(34.8\%S 3T), 
            57.3/\bf{Software}\\(57.3\%S 5T),
            7.9/\bf{Hardware}\\(7.9\%S 1T)
        }
    \end{tikzpicture}
	\caption{Distribution of sentences (S) and topics (T) per topic category}
	\vspace{-2mm}
	\label{fig:topic-cat-dist}
\end{figure}
%




\begin{table}[t]
\caption{Distribution of IoT Security Topics by Total \#of Sentences}
\centering
\resizebox{1\columnwidth}{!}{%
 \begin{tabular}{l l r r} 
 
 \toprule
 Category & Topic Name & \# of Sentences & Distribution(\%) \\ 
 \midrule
 Hardware & IoT Device Config./Set-up & 2402 & 7.9 \\
 \midrule
 \multirow{5}{*}{Software} & Storage/Database Management & 4463 & 14.6 \\
  & Vulnerability/Attack & 3808 & 12.4 \\
  & Framework/SDK-based Security & 3169 & 10.4 \\
  & User/Device Authorization & 3101 & 10.1 \\
  & Crypto/Encryption Support & 3004 & 9.8\\
  \midrule
  \multirow{3}{*}{Network} & Secure Transaction &3855& 12.6\\
  & Secure Connection Config. & 3648& 11.9\\
  & IoT Hub Federation & 3145 & 10.3\\
 \bottomrule
 \end{tabular}%
 }
 \label{iot-topic-distribution}
 
\end{table}

\subsubsection{Results} We found 9 IoT security-related topics, which are grouped into three high-level categories: Software, Network, and Hardware. \fig\ref{fig:topic-cat-dist} shows the distribution of these three high-level categories. The Software category contains the greatest number of topics (5) and covers more than 57\% of all sentences. In Table \ref{iot-topic-distribution}, we show the 9 topics based on the distribution of sentences in IoT security dataset. The distribution column indicates the percentage of sentences in our IoT security dataset that cover this topic. The topic `IoT Device Config' from the Hardware category covers the lowest number of sentences among all topics (2402 sentences, around 7.9\%). It contains discussions of programming and configuration challenges IoT developers face while using Bluetooth/BLE/Microchip interfaces in their IoT devices to setup a service (e.g., setting up a barcode scanner in rpi2 to verify coupon \dq{30721018}). $Q_i$ denotes a question with an ID $i$ in SO. The topic `Storage/Database Management' from the Software category covers the highest number of sentences (4463 sentences, around 14.6\%). This topic contains discussions about developers' problems to database set up, configuration (e.g. \dq{30073381}), and attack like SQL Injection (e.g. \dq{6826176}), Weak Authentication(e.g. \dq{5503812}), Privilege abuse(e.g. \dq{39232652}), and etc.

The other four topics in the software category
are: \begin{inparaenum} \item  Mitigation techniques of
vulnerabilities/attacks in IoT devices (3808 sentences, around 12.4\%), \item Discussions about the vulnerability of specific approaches in frameworks (3169 sentences, around 10.4\%),  
\item Authorization of users and IoT devices (3101 sentences, around 10.1\%), and \item Documentation and usages of Cryptography/Encryption support (3004 sentences, around 9.8\%).
\end{inparaenum}

The Network category has the second highest number of topics (3) and it covers 34.8\% of all sentences. Two of the three topics under the Network category are also among the top four topics (each with around 12\% of sentences). The first two topics are `Secure Transaction' and `Secure Connection Configuration', which contain discussions about ensuring security over secure payments and secure copy/transfer of digital files via/among IoT devices. The last network topic is 'IoT Hub Federation'. Questions in this topic concern the cloud-based setup of IoT devices, e.g., using Azure/AWS/Google cloud. An IoT hub is a managed service in the cloud that acts as a central messaging service to allow bi-directional communication between an IoT application and the smart devices it manages (e.g., connect to MQTT with authenticated Cognito credential \dq{51093154}).

\subsection{RQ$_4$ How do the IoT security topics evolve in SO? }\label{sec:rq-iot-sec-topic-evolve}\label{sec:rq5}
\subsubsection{Motivation} We investigate how the popularity of the three IoT security topic categories has changed over time by determining the total number of new sentences created per topic category over time. A new sentence can only be created if a new post is created. Therefore, the evolution of the topics over time offers us information about the relative popularity of the categories and whether a category is discussed more over time than the other categories.
   
\subsubsection{Approach}For each of the 30K security-related sentences in our dataset, we identify the ID of the SO post from where the sentence is detected. We consider the creation time of the SO post to be the sentence creation time. We group the 30K sentences by the three topic categories. For each category, we compute the total number of sentences created every six months based on the creation time. Following Wan et al.~\cite{Wan-BlockChainTopicSO-IEEETSE2019}, we report the evolution of IoT security topics using the metric Topic Absolute Impact.
Topic absolute impact is calculated by determining the total number of sentences created per the 9 IoT topics every six months since 2008. We do this by first assigning each sentence to its most dominant topic, i.e., the topic with which the sentence shows the highest correlation score. We then assign the creation time to a sentence as the creation time of the post where the sentence is found. Finally, for each topic, we put the sentences into 
time-buckets, where each bucket contains all sentences created within a time window (e.g., from Jan 2009 till June 2009, etc.). We further analyze the absolute growth to correlate the growth in real world contexts for each topic. It is possible that developers might be influenced by the release of new IoT APIs, tools, or devices. As such, we figure out the impact of those real-world contexts in our developer discussion in SO. We first identify some major releases in the IoT domain during the period from 2009 to 2019. Then, we find all the spikes in our absolute growth chart for each topic. Next, we collect all time frames of six months (e.g., January 2010 to July 2010) in which the spikes took place. To identify related real-world events (i.e., major updates and releases), we identify all major updates and releases during that time frame for each topic. Finally, we manually check the availability of discussions for all identified events in our IoT security dataset during that time frame. Based on that, we either count an event as a striking contributor to IoT discussions or discard the event. For example, during the time interval between January 2014 and July 2014, both `vue.js' and `Riot.js' were released. However, we find that our IoT dataset has enough discussion for `Riot.js' only during January 2014 and July 2014. As a result, we discard `vue.js' and count `Riot.js' as a striking event.

\begin{figure}[t]
 \hspace{-0.8cm}
  \includegraphics[scale=.4]{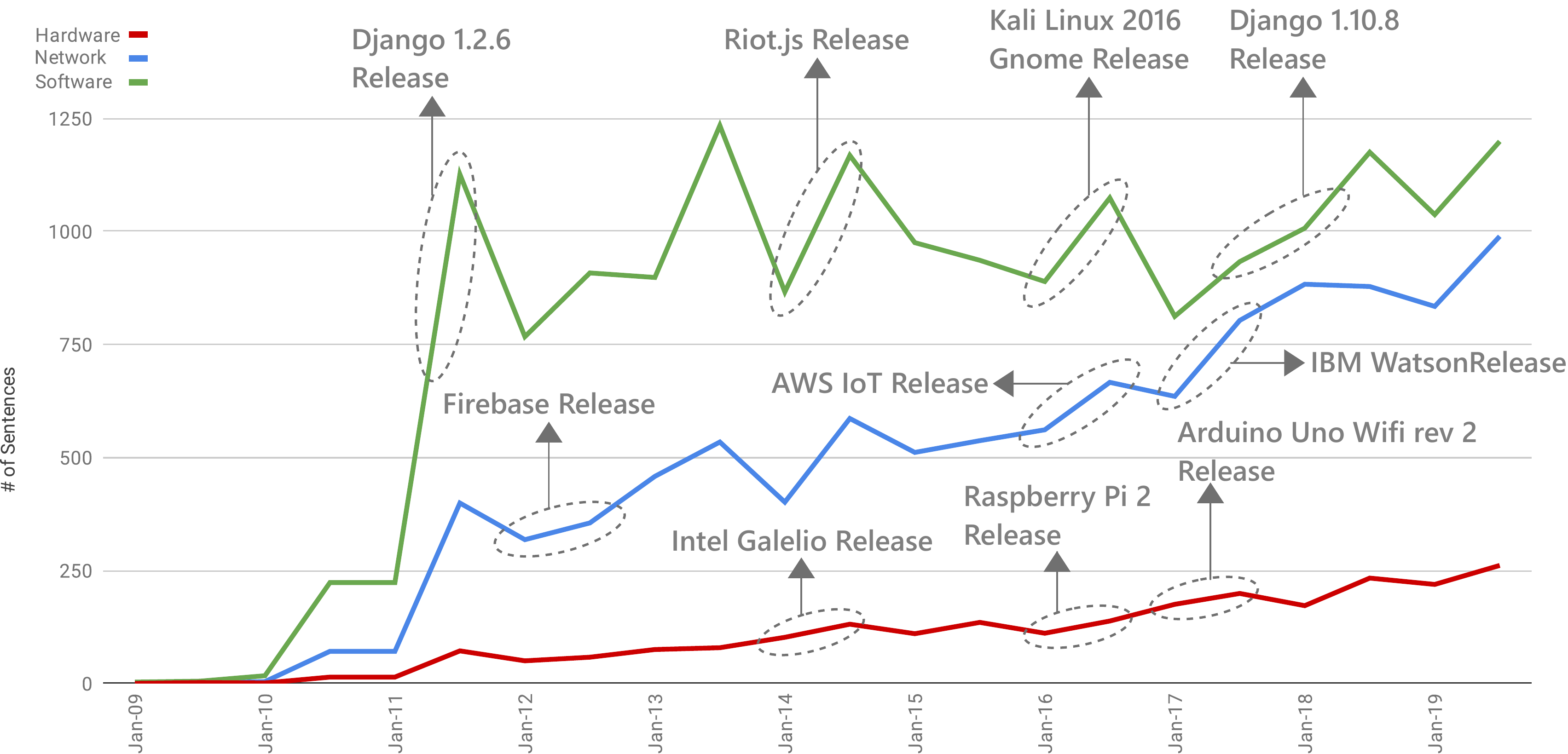}
  \caption{\#New sentences per IoT security topic category each month}

  \label{fig:popularity_overtime}
\end{figure}

%
%

\subsubsection{Results} Figure \ref{fig:popularity_overtime} shows the absolute growth of the three IoT security topic categories over time by showing the number of new sentences per category every six months. While the software category has the most new sentences among the three categories, the network category started to catch up starting from 2017. The number of discussions under Software shows an upswing between January 2011 and July 2011, when Django 1.2.6, a free open-source web framework that encourages rapid development, was launched. Some web security tools (e.g., `XSS', `memory-cache-backed session' and `CSRF btoken') were first introduced in this release. Many discussions in 2014 in our dataset are related to Riot.js. Similarly, the release of Kali Linux in 2006 and Django 1.10.8 gave a spike in software related security discussions. Overall, the periods between June 2011 and June 2014 showcase the rapid adoption of IoT-based SDKs/framekwors (e.g., rpi2 versions). IoT security concerns are raised at the same time due to the emergence of ransomware like CryptoLokcer and banking Trojans. 

%

At the beginning of 2010, the network topic category got high priority due to internet connectivity and cloud storage. A high spike was found at that time. Firebase, a real-time cloud database suited for IoT applications, was released in May 2012. As such, we found many related security discussions across the Network category during this time. Similarly, we found the release of IoT hubs like AWS IoT in late 2015 and IBM Watson IoT in 2017 contributed to many network-related discussions during the respective periods. Recently, the emergence of online transactions and IoT cloud storage solutions has also contributed to the increased popularity of the network category (\#new sentences). 

Hardware related discussions showed continuous growth over the years. Many hardware supported security modules like fingerprint locks, smart doors, biometric devices, etc., have emerged during our study period. We found the release of popular open-source IoT hardware has increased the rate of discussion about that hardware. For example, the release of Intel Galelio back in mid-2012 produced a spike between June 2013 and December 2013. Similarly, after the release of `Rasberry Pi 2' in February 2016 and `Arduino Uno Wifi rev 2' in February 2017, spikes in the curve were visible. 

%
%
%
%
%
%

In conclusion, the evolution of IoT security topic categories and subsequent topics can provide insight into new, emerging, or existing threats.

%
%
%

\section{Discussion}\label{sec:discussion}
For our experiments, we use SO data dumps up to September 2019. However, we also
investigate the most recent data dumps of SO. We use the online Stack Exchange data
\href{https://archive.org/details/stackexchange}{archive} to get the most recent
SO data dump, which was for January 2022. We collect IoT datasets from this data dump by
following all the steps we described in \sec \ref{sec:iot-data-collection}. 
We found that many SO posts related to IoT were deleted by SO in the January 2022 data dump that were previously
available in the September 2019 data dump. As such, we first analyze the extent of 
such deleted posts in the January 2022 dump (in \sec\ref{sec:case-deleted}). 
We then compare the results of our empirical studies for RQ3 (recall \sec\ref{sec:rq4}) and RQ4 (recall \sec\ref{sec:rq5}) between the two data dumps, i.e., Sept 2019 and Jan 2022 in 
\secs\ref{sec:prevalence-after-2019}, \ref{sec:evolution-after-2019}.

\subsection{The ``Case'' of Post Deletion by Stack Overflow}\label{sec:case-deleted}
As per the data dumps of September 2019, there are around 53K IoT related posts,
whereas the most recent data dumps (January 2022) have only 45K IoT related
posts. This indicates that most recent data dumps have fewer IoT-related posts
even though our trend charts from \fig\ref{fig:popularity_overtime} based on Sept 2019 data dump showed 
that IoT-related discussions have an upward trend in recent times. We
investigate this drop in the number of posts in Jan 2022 data dumps. We identified
all the missing posts that are available in the September 2019 data dumps. Then,
we check the recent activity and status of these missing posts. We find that
these posts were deleted in between September 2019 and Jan 2022. As a
result, these posts appear in the 2019 data dumps but do not appear in recent
data dumps. For example, `ESP32' tagged SO question \dq{56098708} was created in
May 2019, but later it was deleted. We further apply Secbot to the new IoT data
and find only 6K security-related sentences. Compared to the 30K
security-related discussion in RQ$_3$(\sec \ref{sec:rq4}), these results are
quite low. This is because the size of the dataset is smaller and most of the
posts related to the security discussions are removed. For example, an
encryption-related post, \dq{11501569}, that contains multiple security-related
discussions, has been removed in recent data dumps.
\begin{figure}[t]
 \hspace{-0.8cm}
 \centering
  \includegraphics[scale=.5]{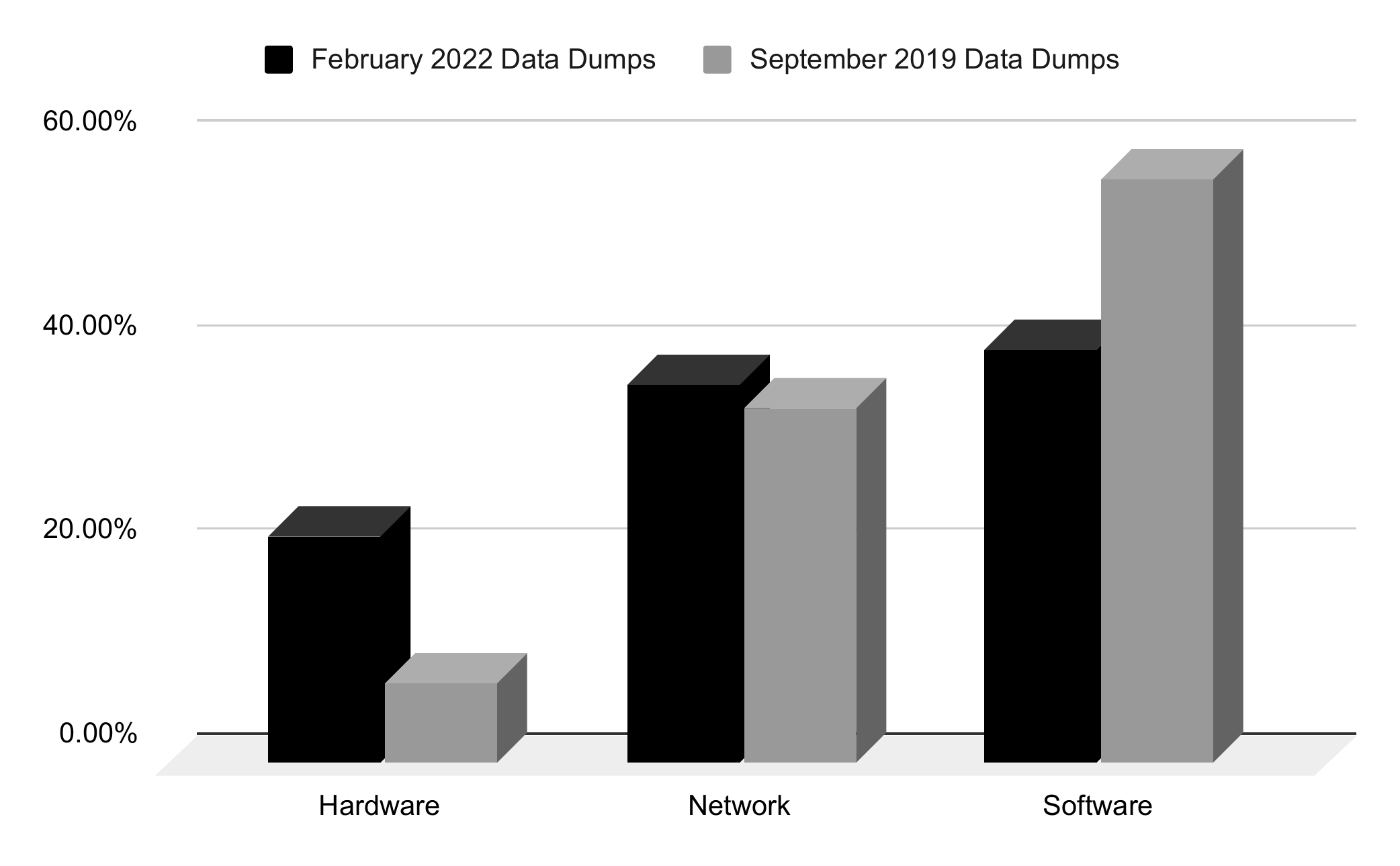}
  \caption{Topic distribution in February 2022 SO dumps and September 2019 SO dumps}

  \label{fig:compare-distribution}
\end{figure}

\subsection{Prevalence of IoT Security Topics After 2019}\label{sec:prevalence-after-2019}
Following the RQ\ref{sec:rq4}, we apply topic modeling on 6k sentences we got
from the Jan 2022 data dump. We find that the optimal number of topics for this
dataset is 6. Then we manually label these six topics. The topics are: `IoT
Device Config/Set-up', `Secure Connection', `Secure Transaction',
`Framework/SDK-based Security', `Authentication Management', and
`Crypto/Encryption Support'. We further investigate how these topics differ from
previously found topics in RQ$_3$ (\sec \ref{sec:rq4}). Although five topics are
present in our previous findings, `Authentication Management' is different from
what we found previously in `User/Device Authentication'. `Authentication
Management' is comprised of discussion related to authentication mechanisms,
devices, techniques, protocols, implementation supports, etc., whereas
`User/Device Authentication' only consists of user or device level
authentication. On the contrary, we didn't find any discussion related to
`Storage/Database Management', `Vulnerability/Attacks', or `IoT Hub Federation'.
This happens because of the removed posts in recent data dumps. We further
categorize these topics into Software, Hardware, and Network group. We find 3
software-related topics (i.e., `Framework/SDK-based Security',
`Crypto/Encryption Support', and `Authentication Management'), 2 network-related
topics (i.e., `Secure Connection' and `Secure Transaction' ), and a hardware
related topic (i.e., `IoT Device Config/Set-up'). There are 22.3\% hardware,
37.2\% network, and 40.5\% software related security-discussions in the February
2022 data dumps. Finally, we compare the topic distribution in the most recent
data dumps (i.e., February 2022) and the September 2019 data dumps. We show the
comparison in \fig \ref{fig:compare-distribution}. The relative percentage of
hardware-related discussions is higher in February 2022 data dumps than in
September 2019 data dumps. Although the network topic holds similar percentages
in both data dumps, the software topic has a lower value in the February 2022
data dumps. This is because most of the software-security related IoT posts were
removed in the February 2022 data dumps, and in recent times, developers are
discussing hardware-security more than they did in past times.
\begin{figure}[t]
 \hspace{-0.8cm}
 \centering
  \includegraphics[scale=.65]{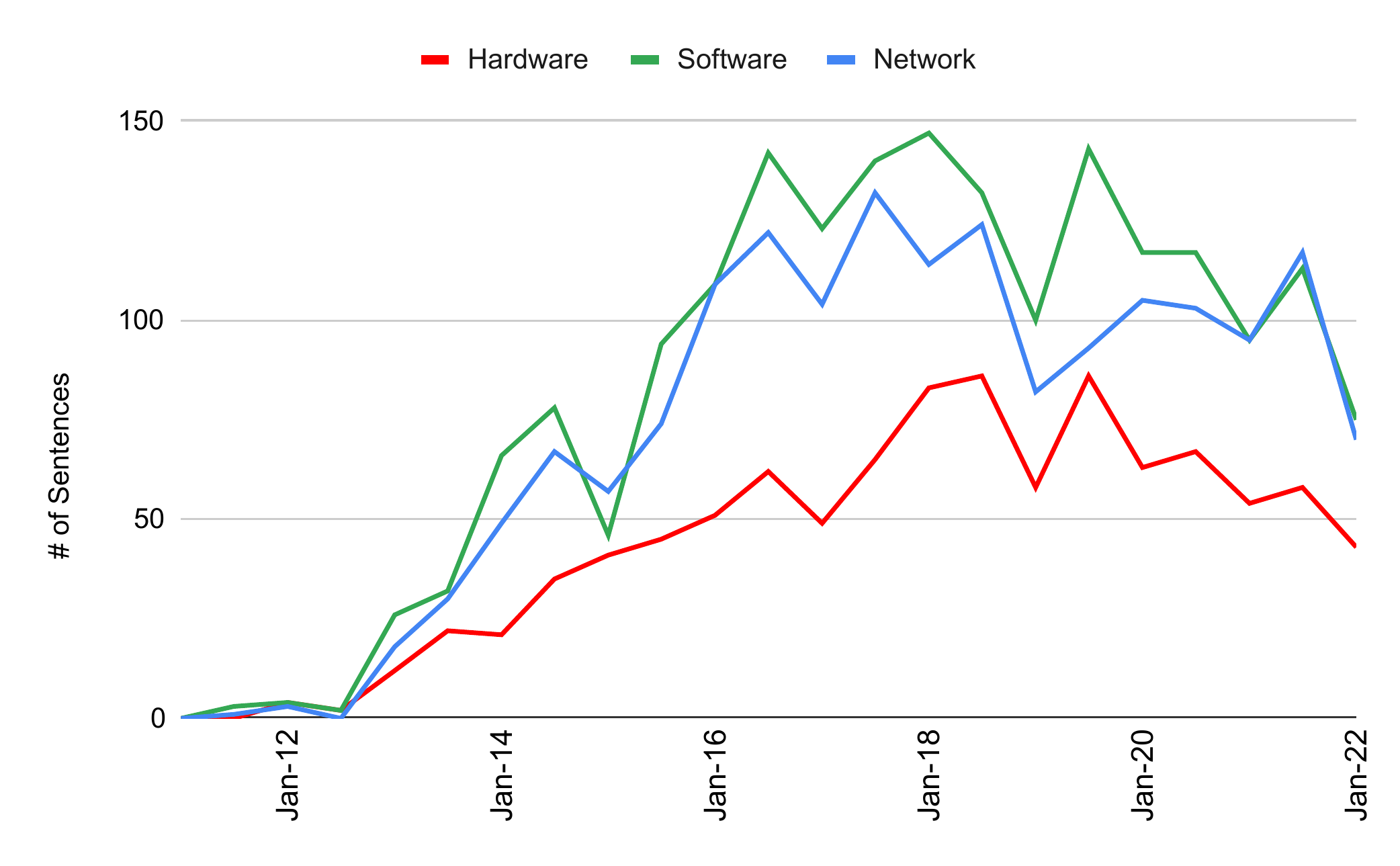}
  \caption{Absolute growth of each IoT security topic category in each month for most recent data dumps (February 2022)}

  \label{fig:trend-chart-new-iotdata}
\end{figure}

\subsection{Evolution of IoT Security Topics}\label{sec:evolution-after-2019}
We analyze how the topics found in the Jan 2022 data dump evolved over the time period. We first calculate the
absolute growth of the topics following RQ$_4$ (\sec \ref{sec:rq5}). Then we
generate a trend chart showing how the three topics (i.e., Hardware, Software,
and Network) evolve over the time period between 2011 and 2022. We report the
trend chart in \fig \ref{fig:trend-chart-new-iotdata}. From this chart, we find
that security related discussions follow an upward trend for all three topics
(i.e., hardware, software, and network) before 2019, similar to our previous
findings in \sec \ref{sec:rq5}. However, there is a downward slope after 2019
for all three topics. One of the possible reasons for this downward swing could
be the lack of releases of any software, hardware, and network related devices,
tools, technologies, etc. during this period. Another contradictory finding is
that network and software related discussions have almost similar numbers, but
our previous finding says there are more software-related discussions than
network ones. This could be because of the removed posts in the February 2022
dumps.

\subsection{Discoverability of IoT security sentences}\label{sec:discoverability-iot-security}

\begin{figure}[t]
    \centering
	\begin{tikzpicture}[scale=0.45]-
    \pie[
        /tikz/every pin/.style={align=center},
        text=pin, number in legend,
        explode=0.0, rotate=100,
        color={black!60, black!0},
        ]
        {
            8/\bf{8\% Found}\\out of 30K\\sentences using\\frequently used 10 security tags,
            92/\bf{92\% Missing}
        }
    \end{tikzpicture}
    \caption{IoT security sentences dicoverability via frequently used SO security tags}

	\vspace{-2mm}
	\label{fig:hit-miss-posts}
\end{figure}
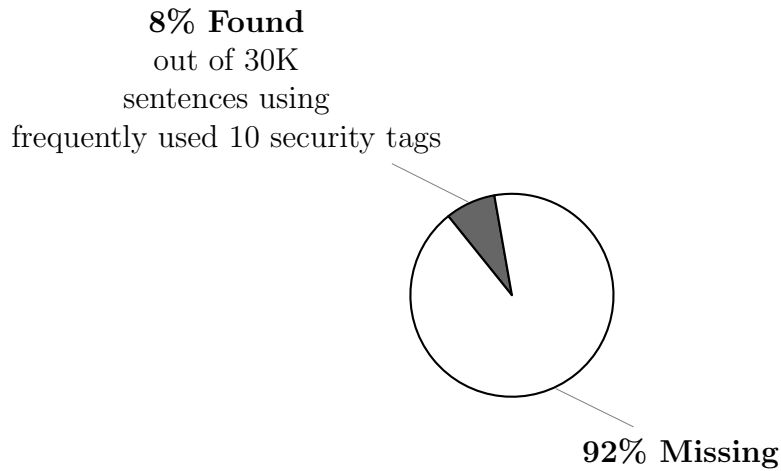


Our study data is collected from SO, by identifying questions in SO labeled as one of the 75 IoT related tags. As we noted in  \sec\ref{sec:iot-data-collection}, the 75 tags were previously used to learn IoT developer discussions in SO by Uddin et al.~\cite{uddin-iot}, who found that the most popular and most frequently discussed topics were related to the adoption of security measures in IoT devices (e.g., secure messaging). In addition, the same dataset was also used by Uddin~\cite{Uddin-iot-SERP4IoT2021} to determine the prevalence of security-related discussions at the sentence level. Both studies show that the dataset contains diverse discussions about security for IoT. Note that not all the 75 IoT tags have explicit reference to any security measures, but we observed that many of the questions (and their answers) labeled as the tags contained security discussions. In this section, we offer quantitative evidence on the extent of security discussions in questions labeled as non-security tags as follows.

\rev{We investigate whether SO security-specific tags cover all IoT security sentences. Thus, we pick 10 frequently used security-related tags, i.e., `security', `ssh', `ssl', `passwords', `authentication', `authorization', `encryption', `cryptography', and `hash' in SO. Then, we collect all posts from our 53K IoT posts with at least one of the tags. In the collected posts, we only manage to find 8\% of 30K sentences, i.e., 92\% of our IoT security sentences are not discoverable using the 10 tags (see \fig \ref{fig:hit-miss-posts}). We also investigate the missing posts in the latest data dumps (January 2022) using these 10 security tags. The result shows the numbers are similar, only 10\% posts can be discovered with these 10 tags in 2022 data dumps. This indicates that SO security tags do not cover all security discussions. However, our security discussion includes sentences that are tagged with security tags.}    
\section{Implications of Findings}\label{sec:implications}
The findings from our study can guide the following major stakeholders in the rapidly emerging IoT ecosystems: \begin{enumerate}
\item \bf{IoT Security Enthusiasts} to learn about IoT security aspects from IoT developer discussions, 
\item \bf{IoT Vendors} to offer tools and techniques to support security in IoT devices,
\item \bf{IoT Developers} to determine the current trends in IoT security based on developer discussions and to use the information to guide future development needs, 
\item \bf{IoT Educators} to develop tutorials and documentation to educate security principles to IoT practitioners, and
\item \bf{IoT Researchers} to improve the detection of IoT security aspect in developer discussions and to study how to properly address the difficulty in the adoption of security practices for IoT devices and techniques. 
\end{enumerate} We discuss the implications below.
\begin{figure}[t]
  \centering
  \includegraphics[scale=.55]{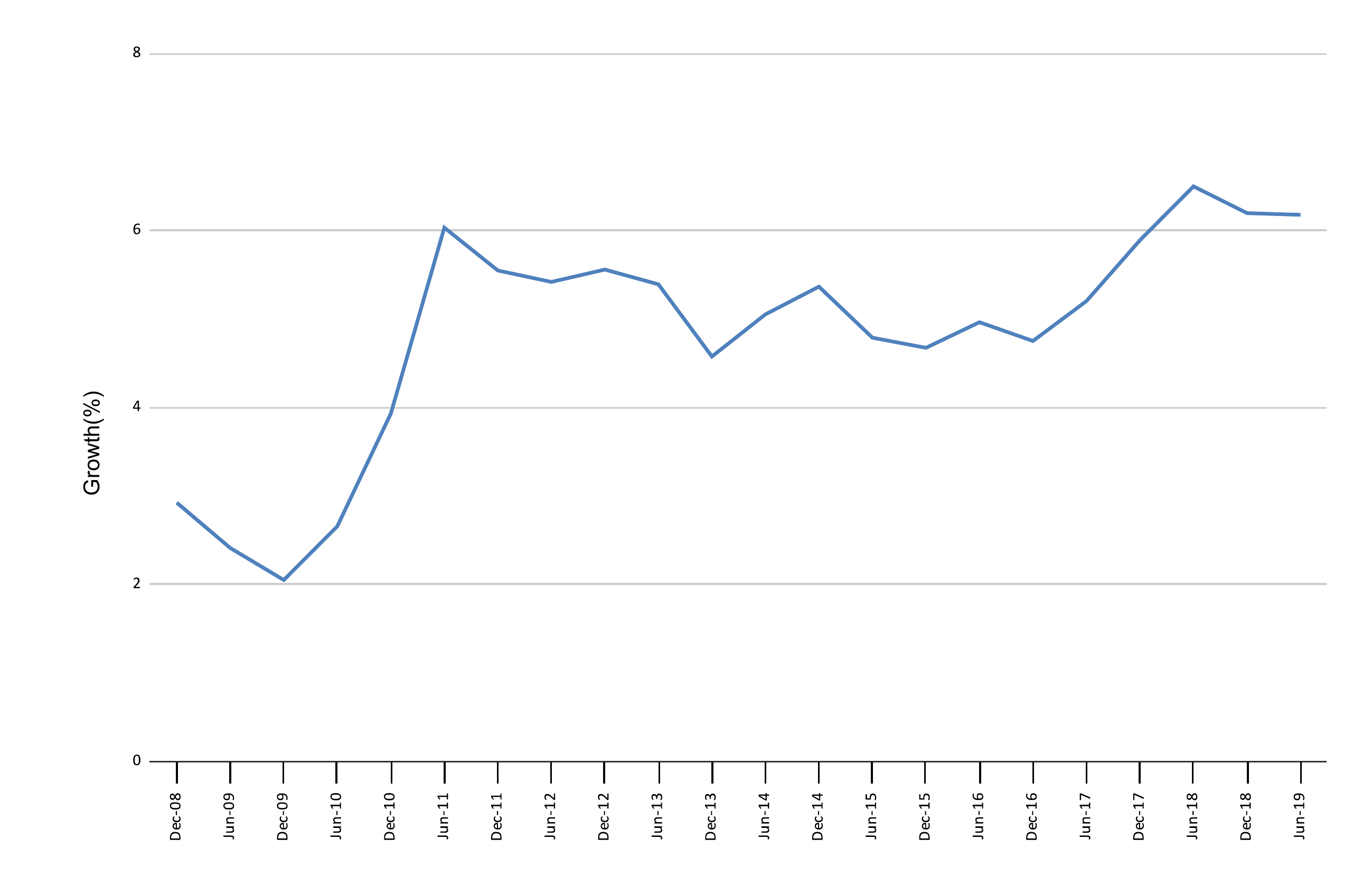}

  \caption{Relative Growth of security related sentences over time}

  \label{fig:relative_popularity}
\vspace{-3mm}
\end{figure}

\bf{\ul{IoT Security Enthusiasts.}} IoT enthusiasts can be IoT practitioners or
simply IoT market watchers, but they nevertheless form a crucial part of the IoT
ecosystem in various ways (e.g., buying/analyzing IoT devices/solutions, etc.).
The development and adoption of IoT-based solutions by developers is facilitated
by the exponential growth of IoT devices, software, and platforms. Our
increasingly interconnected digital world relies on smart devices built using
the IoT, which means security for IoT devices is paramount and it is the job of
IoT developers to adopt proper security measures for their solutions. Indeed, if
we compare the number of new security sentences per month to all sentences in
the month in our 53K SO posts, we observe a slow but relative growth of IoT
security-related discussions, especially after 2016 (see \fig
\ref{fig:relative_popularity}). This means that interest in IoT security is
increasing among IoT developers in SO. Therefore, IoT security enthusiasts can
benefit from the security discussions by IoT developers in SO. As we noted in
\sec \ref{sec:intro}  (\fig \ref{fig:motivation-post-sentence}), security
discussions in SO IoT posts can be buried inside general or non-security IoT
discussions. Therefore, to offer security-specific insights from SO IoT
discussions, we need tools to automatically detect those. The high performance
of our developed tool, SecBot, is suited to meet the needs.


\begin{sidewaysfigure}
    \includegraphics[scale=1.1]{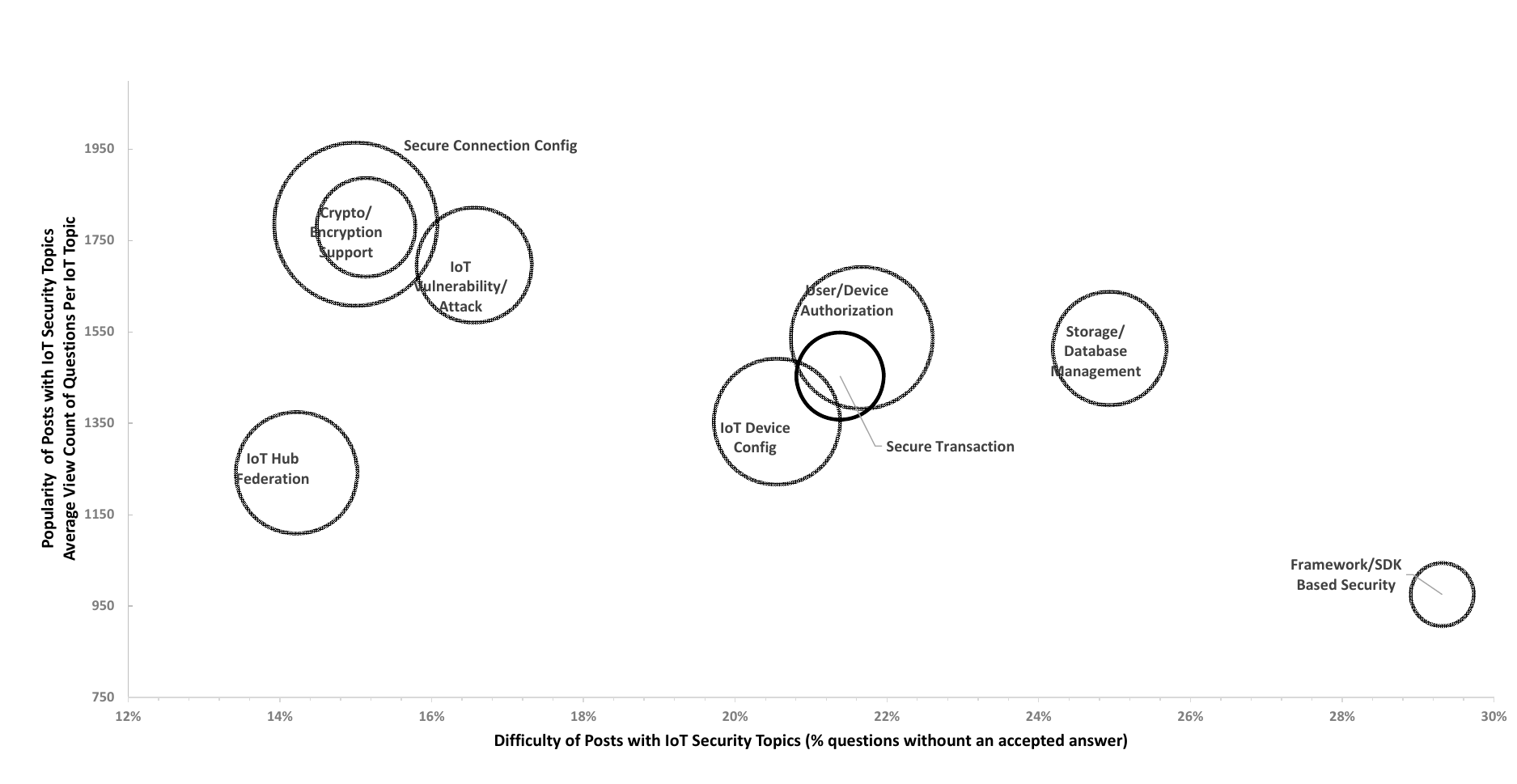}
    \caption{Tradeoff between popularity and difficulty of IoT security topics in our IoT dataset}
    \label{fig:bubble}
\end{sidewaysfigure}
\bf{\ul{IoT Vendors.}}  IoT vendors need to support IoT developers with proper and usable secure IoT techniques. To determine what is appropriate and what is working, though, the vendors need to know the problems faced by IoT developers. Forums like SO have become a go-to place to look for solutions to their technical problems. Thus it has become essential for the IoT vendors to keep an eye on this open discussion forum. Our research can help them in two ways. \textit{First,} the high precision and recall of our tool SecBot will help IoT vendors reliably find security-related issues in SO. They can analyse those discussions and take a decision accordingly. \textit{Second,} our findings from RQ$_4$ (\sec \ref{sec:rq5}) show that IoT developers are aware of major releases and immediately discuss them more frequently. Thus, we can say the vendors can get quick feedback from the developers about their most recent release of any IoT device, tools, techniques, or software. This could play a vital role in their upcoming releases. For example, if there are multiple bugs or issues, they can fix those bugs in their next release or offer a quick release fixing those bugs. Overall, such insights can be crucial for vendors to improve their offerings and to compare the solutions of their competitors. Hence, they could be well ahead of their competitors.  

In \fig \ref{fig:bubble}, we show a bubble chart to show the tradeoff between the popualrity vs difficulty of IoT security posts. The chart is 
constructed as follows. Out of the 30K IoT security sentences in our dataset, we pick the sentences that we find in the SO question posts. 
For each of the 9 IoT security topics, we pick the sentences that we find in the SO question posts. For the total  number of distinct questions thus 
found for a given IoT security topic, we analyze two metrics as follows:
\begin{inparaenum}[(1)]
\item Popularity analysis based on the average view counts of the associated questions to the topic.
\item Difficulty analysis to get an accepted answer to a question by computing the percentage of question per topic that remain without an accepted answer.
\end{inparaenum} Both of the metrics have been used previously in literature to determine the popularity and difficulty of topics generated from SO posts. 
The x-axis in \fig \ref{fig:bubble} shows the difficulty score of each IoT security topic, and the y-axis shows the popularity score of the IoT security posts. The size of each topic bubble is based on the number of distinct questions assigned to the topic. Therefore, the more right a topic in the chart, the more difficult it is in SO. Topics like `Framework/SDK Based Security', and `Storage/Database Management', and `IoT Device Config' are among the most difficult. The topic `IoT Device Config' is among the most difficult as well as the most popular IoT topics. Therefore, IoT vendors can accelerate the adoption of security in IoT Microschip and Bluetooth (e.g., BLE) devices, which will then increase the chance of getting answers to the problems. 

\bf{\ul{IoT Developers.}} The security of IoT devices and solutions is of paramount importance. As the hacker community is looking for the single opportunity to breach the IoT system or device, the developers should not consider ignoring any corner of security vulnerability which can be exploited. We thus need IoT developers to stay aware of the IoT security trends and to adopt prevalent IoT tools. The developer community is one of the best places that could help them keep pace with the cutting edge technologies that can't be breached, at least for that time frame. Such insights can benefit from such precise, specific, but automatically mined security knowledge to stay aware of the trends related to IoT security. In addition, such knowledge can also help them make better decisions, like picking a popular IoT security technique over another. For example, in \fig \ref{fig:bubble}, we find that the topic `IoT Hub Fedearation' is the least difficult, while also not very popular. This means that IoT developers leveraging cloud infrastructure to implement security methods for their IoT devices are not finding enough support from SO developers. The topic `Secure Transaction' between IoT devices or between IoT and non-IoT devices contain a large number of questions, which are also popular. This means that IoT developers can devote time to learn and implement secure transactions using their IoT devices, and they can also inquire with other IoT developers in SO about the problems they face.

The developer community cannot rely on new technologies every time, as new technologies can be challenging, complex, or incompatible with the current environment. For example, a developer asks how he can create a digital signature function PDF from Hardware Security Modules (HSM) using C\# in SO question \dq{54922093}. However, the discussion section suggests that he can't use HSM directly in C\# instead he needs to go through `CAPI' or `CNG' api or `ncryptoki'. Developers could face these challenges in the middle of any project, which could cost more time or, at worst, they could end up switching to another framework or tool. This scenario can be avoided if developers gather the necessary information about the required apis, devices, or software. Developer forum like SO is the ideal place to collect those information. Our research provides information about IoT security topics. Further mining of their discussed topics such as which features they are discussing, what are the sentiment of those discussion, etc. will be helpful for everyone to learn about an api, tools, frameworks, software, or devices. From this informative feedback, the developer community can decide whether they adopt the new technology or not. Thus, they can select their developing tools and frameworks wisely and minimize the risk of a project.

\bf{\ul{IoT Security Educators.}} Our research findings are more promising for IoT educators in many ways. From RQ$_4$ (\sec \ref{sec:rq5}), we learn that IoT topics are getting more attention and thus the responsibility of the IoT educators is also getting escalated. As the IoT security demands are increasing, the educator must prepare quality contents to motivate the newcomers and to enlight current practitioners.  Thus, to create content, they can follow the topics we find in RQ$_3$ (\sec \ref{sec:rq4}). They should also be quite selective in selecting the most recent techniques and tools over the old ones. Otherwise, the learners will lack the new tools, which will eventually end up going backward instead of forward. SO can be a helping hand here for them. As SO contains IoT security discussions about all those 9 topics, they can collect that data to analyse and prepare their lessons. Moreover, as we previously presented, one can find enough discussions related to the most recent releases of any IoT security tools, software, and devices. The educator can also provide these insights regarding the most recent IoT tools, software, and devices into their lesson. This will also be beneficial for the learner.

IoT security educators can use the bubble chart in \fig \ref{fig:bubble} to prioritize their efforts to develop security tutorials and documentation for IoT developers and practitioners. For example, the topic `Secure Connection Config' has the highest number of questions. It is also among the most popular topics, while also less difficult than most topics in terms of getting an accepted answer. Therefore, IoT security educators can analyze the questions related to user/device authorization in SO to develop comprehensive tutorials for developers. The topic `Crypto/Encryption Support' has the second most questions, while it is the most popular yet one of the most difficult topics. Therefore, the IoT security educators can analyze the questions in SO and consult IoT resources to ensure that IoT developers can learn about the risks associated with leaving IoT ports open and how they can configure them properly to ensure security.

\bf{\ul{IoT Researchers}} 
This research creates a vast space for the researcher to go forward in multiple dimensions. Security aspect detection is a major task that we accomplish in this paper. However, this task requires more attention in the future. We find our Secbot has a maximum performance of 0.935, which seems like a good result considering no prior security detector has achieved that far. But when we figure out the errors made by our Secbot, we find multiple implicit contexts, ambiguous contexts, and ambiguous keywords related errors. These errors can further be researched to extract informative information like the identification of key phrases or words that cause ambiguity. Individual research can be conducted on how these key words or phases can be detected automatically. In another way, these error categories give us a way of improving the performance of Secbot. The researcher can approach this in three ways. \textit{First,} they can add IoT contexts to Secbot. For example, they can use some rules based patterns to extract information and then feed it to the Secbot. \textit{Second,} they can add more samples to reduce the sparsity of security-aspects in the dataset. For this, one may consider only adding the samples that are close to those error category types. They can even try to make a balanced dataset to see how the performance changes. \textit{Third,} they can try a deep model that can only discern samples related to those error categories, and then they can apply ensemble Secbot and the deep model.

Besides this, we find that domain-specific BERTOverflow fails to perform as well as the generic model Secbot. Additionally, the correlation coefficient score from Table \ref{correlation} suggests that the results of BERTOverflow vary from the results of Secbot. This could be an interesting topic to delve deeply into. The researcher can explore the misclassification made by BERTOverflow to compare the results between Secbot and BERTOverflow. If there is a significant difference in error categories between these two models, they can consider developing an ensemble model. Another direction of research exploration could be developing a purely domain-based security aspect detector. Although BERTOverflow has knowledge about StackOverflow, compared to its overall knowledge, SO knowledge is not sufficient to represent it as an intelligent SO expert. Thus, an entirely SO knowledge-based model may be more suitable in security-aspect detection.

IoT researchers can analyze the security discussions to learn about the specific challenges that IoT developers are facing based on their real-world experience. Such insights can be useful for researchers to invent new techniques and tools for IoT security. It is important that research in IoT security can be influenced by the emerging trends in IoT security. As we observed in \sec \ref{sec:rq-iot-sec-topic-evolve} (\fig \ref{fig:popularity_overtime}), we see an almost equal number of sentences per the the two topic categories (Software and Network) starting from January 2017. This means that security research in the IoT needs to put equal emphasis on both software and network security. Similarly, discussions about secured hardware for IoT devices are also increasing over time, although not as much as the software and network topics. IoT security researchers need to design and develop innovative techniques to secure IoT software and networks. One of the 9 topics is `Vulnerability/Attack Concerns' in IoT devices, which points to the issues IoT developers are facing with regards to addressing specific vulnerabilities in their IoT-based solutions. As we find in \fig \ref{fig:bubble}, this topic is also considerably popular and difficult. Therefore, IoT software and hardware security researchers can use the discussions to develop tools and techniques. Data management and storage are subject to the topic `Secure Data Management', which can benefit from research in database security.

Additionally, the research community can conduct rigorous research on IoT security topics such as what types of questions developers asked in each topic category, how the new release affects the developers, what factors developers discussed most about any new release, etc.

\section{Threats to Validity}\label{sec:threats}
\bf{Internal validity} threats relate to the authors' bias while conducting the analysis. We mitigated the bias in our benchmark creation process and topic labeling processes by computing agreements (security sentences) and labeling together (topic modeling). During the topic labeling, the authors communicated over Skype and using Google drive to reduce individual biases. The agreement between the two coders is above 95\% all the time. We use a standard random sampling technique to reduce locality biases in our benchmark dataset. The machine learning models are trained, tested, and reported using standard practices. There was no common data between the training and test sets. We shuffle the training and testing datasets in each iteration to introduce a new scenario every time. We perform cross-validation on the entire dataset to make sure our assessment is not biased towards any subsets. We follow standard hyperparameter tuning to minimize the effects of hyperparameters on the final results.

\bf{Construct validity}  threats relate to the selection and creation of IoT security dataset. As we include all tags related to IoT tags, security-related tags are also present there. However, there may be some security tags that have not been included in the dataset creation process. In the future, SO may include more IoT and security tags, which could challenge our dataset. However, as we discussed in \sec\ref{sec:discoverability-iot-security}, we observed that IoT developers in SO did not constrain their security-related discussions in SO to only questions labeled as security-related tags. In fact, more than 90\% of our security-related sentences are not covered by the security-specific tags in our dataset. As such, in \sec\ref{sec:implications}, we discussed the implications of our automated tool to detect security-related sentences in SO, which could find such sentences even in SO IoT tags that do not explicitly refer to any security issues. Beside this, construction validity threats relate to the difficulty of finding data to create our IoT security-related sentences. Our benchmark creation process was exhaustive, as we processed more than 53K posts from SO. The evolution of IoT topic categories considers post-creation time as sentence creation time. A post can be edited after its creation with new sentences, but we observed less than 1\% of such cases in our dataset. We find security-specific 30K sentences out of 672K sentences in the IoT dataset using the best performing SecBot model. During the topic modeling phase, we observed only a few non-security related sentences. Further investigation can be carried on. This may give us more insights about the model, which we leave as our future task. In addition, we select a topic for each sentence based on the higher coherence score of the topics. During the topic labeling step, we label each topic by analyzing the topic top 30 words and posts associated with the label. This approach is consistent with the related work that also analyzed topics in SO posts~\cite{Bagherzadeh-BigdataTopic-FSE2019,abdellatifchallenges}.

\bf{External validity} threats relate to the generalizability of our findings. Our developed model, SecBot, demonstrates high accuracy. As such, the model is expected to offer good performance for other developer forums. There is a possibility that IoT developers may discuss different types of security discussions in other forums. For example, we found that most of the IoT developers discuss Database/Storage related security in SO. It is possible that in other forums, they discuss more about encryption. A detailed evaluation of SecBot in other forum data is our future work. In summary, our model exhibits good performance in this research direction, but the results should not be taken as an automatic implication of the same result in general. An extensive analysis of the diverse nature of challenges and characteristics can validate the transposition of the results to other domains. Besides this, SecBot is designed to work on textual datasets. However, SO contains code-snippets, logging, and urls which are filtered out during our dataset creation. Therefore, if SecBot is applied to SO posts containing source codes, logging, or urls, the model's performance may drop.

\section{Related Work} \label{sec:background}
Related work can broadly be divided into \bf{Studies} to understand and \bf{Techniques} to detect/mitigate IoT security issues.

\nd\bf{\ul{Studies.}} Literature in IoT so far has focused on
surveys of IoT techniques and
architectures~\cite{Sethi-IoTArchitecture-JECE2017,Fuqaha-IoTSurveyTechnologiesApplications-IEEECST2015},
the underlying middleware solutions (e.g.,
Hub)~\cite{Chaqfeh-ChallengesMiddlewareIoT-2012}, the use of big data analytics
to make smarter devices~\cite{Marjani-IoTDataAnalytics-IEEEAccess2017}, the
design of secure protocols and
techniques~\cite{Fuqaha-IoTSurveyTechnologiesApplications-IEEECST2015,Khan-IoTSecurityReview-FGCS2018,Zhang-IoTSecurityChallenge-SOCA2014}
and their applications on diverse domains (e.g.,
eHealth~\cite{Minoli-IoTSecurityForEHealth-CHASE2017}), the Industrial adoption
of
IoT~\cite{Liao-IndustrialIoT-IEEEIoT2018},
and the evolution and visions related to IoT
technologies~\cite{Pretz-TheNextEvolutionInternet-IEEEMagazie2013,Sharma-HistoryIoT-ElsevierIoT2019}.
The unauthorized inference of sensitive information from/among IoT devices 
is a prevalent concern\cite{Celik-IoTSensitiveInformationTracking-USENIX2018}.

We are aware of no previous research that focused on understanding
IoT security discussions in SO.  
In SE, topic modeling is used to learn aspects like software logging~\cite{Li-StudySoftwareLoggingUsingTopic-EMSE2018}, 
feature location~\cite{Cleary-ConceptLocationTopic-EMSE2009,Poshyvanyk-FeatureLocationTopic-TSE2007},
traceability linking~\cite{Rao-TraceabilityBugTopic-MSR2011,AsuncionTylor-TopicModelingTraceabilityWithLDA-ICSE2010a},
software and source code evolution~\cite{Hu-EvolutionDynamicTopic-SANER2015,Thomas-SoftwareEvolutionUsingTopic-SCP2014,Thomas-EvolutionSourceCodeHistoryTopic-MSR2011},
source code categorization~\cite{Tian-SoftwareCategorizeTopic-MSR2009}, code refactoring~\cite{Bavota-RefactoringTopic-TSE2014}, 
defect analysis~\cite{Chen-SoftwareDefectTopic-MSR2012}, and various software maintenance
tasks~\cite{Sun-SoftwareMaintenanceTopic-IST2015,Sun-SoftwareMaintenanceHistoryTopic-CIS2015}.   
The SO posts are subject to topic modeling to understand concurrency~\cite{Ahmed-ConcurrencyTopic-ESEM2018}, big
data~\cite{Bagherzadeh2019} and chatbot issues~\cite{abdellatifchallenges}. 
Yang et al.~\cite{yang2016security} studied security-topics in SO. They used SO tags to identify mobile security which might miss many security discussions which were not tagged as any security related tags. Our studies resolve this issue by following Uddin et al. \cite{uddin-iot} to collect all IoT posts. Next, they applied topic modeling to the security posts. As the recent studies have found that larger documents may have multiple topics, LDA topic modeling is unstable for such cases \cite{LDA-challenge}. We thus use sentence level topic modeling. We develop a precise security detector, SecBot, to identify security related sentences from SO posts and apply LDA topic modeling on our security dataset. Unlike  Yang et al.~\cite{yang2016security}, we focus on IoT security topics.

\nd\bf{\ul{Techniques.}} IoT devices 
can be easy target for cyber threats~\cite{Zhang-IoTSecurityChallenge-SOCA2014,Frustaci-IoTSecurityEvaluation-IEEEIoTJournal2017}. As such, significant research efforts are underway to improve IoT security. Automated IoT security and safety measures are
studied in Soteria \cite{Celik-IoTSafetySecurityAnalysis-USENIX2018},
IoTGuard~\cite{Celik-IoTDynamicEnforcementOfSecurity-NDSS2019}. Encryption and
hashing technologies make communication more secure and certified~\cite{Tedeschi-LikeSecureIoTCommunications-IEEEIoT2020}. Many authorization
techniques for IoT are proposed like SmartAuth~\cite{YuanTian-APIBot-ASE2017}. For smart home security, IoT security techniques are proposed like Piano~\cite{Gong-IoTPIANO-ICDCS2017},  smart authentication~\cite{He-RethinkIoTAccessControl-USENIX2018}, and cross-App Interference threat mitigation~\cite{Chi-SmartHomeCrossAppInference-DSN2020}. Session management and token verification are used in web security to
prevent intruder getting information. Attacks on Zigbee, an IEEE specification used to support interoperability can make IoT devices vulnerable~\cite{Ronen-IoTNuclearZigbeeChainReaction-SP2017}. Smart gateway for IoT is proposed to tackle
malicious attack~ \cite{Hussain-IoTSecurityLink-DCOSS2019}. To the best of our knowledge, our developed SecBot+ is the first DL model that can be used to  
automatically detect IoT security-related developers discussions. IoT researchers can gain insights to 
offer increased support/security for a problematic IoT device as observed in the developer discussions. 

%
\section{Conclusions} \label{sec:conclusion}
The rapid adoption of IoT-based solutions has raised concerns about the security of IoT devices and communications. As such, it is important to understand the problems developers discuss when discussing their usage of IoT security tools and techniques in online technical forums like SO. With a view to automatically detecting developers' security discussions at the granularity level of sentences, in this paper, we have investigated a total of five advanced pre-trained language-based deep learning techniques (e.g., BERT, RoBERTa). The best-performing model, SectBot, based on RoBERTa, offers an F1-score of 0.935 to detect IoT security discussions. We use SecBot to automatically mine all the 30K IoT security-related sentences from the 53K IoT posts from SO. We apply topic modeling to the 30K sentences to find IoT security topics in developer discussions. We observed 9 topics that are grouped into three categories: Software, Network, and Hardware. Our developed tools and study findings can guide automated collection and analysis of IoT security problems in developer discussions.


\begin{small}
\bibliographystyle{abbrv}
\bibliography{consolidated}
\end{small}

\end{document}